\documentclass{article}

\usepackage{microtype}
\usepackage{graphicx}
\usepackage{subcaption}
\usepackage{booktabs}
\usepackage{hyperref}
\usepackage{multirow}
\usepackage{enumitem}
\usepackage{gensymb}
\usepackage{fancyvrb}

\usepackage[accepted]{icml2026_GenAICreativity}
\usepackage{amsmath}
\usepackage{amssymb}
\usepackage{mathtools}
\usepackage{amsthm}
\usepackage{fvextra}

\usepackage[capitalize,noabbrev]{cleveref}

\theoremstyle{plain}

\theoremstyle{definition}

\theoremstyle{remark}

\icmltitlerunning{Evaluating Design Video Generation: Metrics for Compositional Fidelity}

\begin{document}

\twocolumn[
  \icmltitle{Evaluating Design Video Generation: \\
    Metrics for Compositional Fidelity}

  \icmlsetsymbol{equal}{*}

  \begin{icmlauthorlist}
    \icmlauthor{Adrienne Deganutti}{yyy}
    \icmlauthor{Dingning Cao}{yyy}
    \icmlauthor{Jaejung Seol}{yyy}
    \icmlauthor{Elad Hirsch}{yyy}
    \icmlauthor{Purvanshi Mehta}{yyy}
  \end{icmlauthorlist}

  \icmlaffiliation{yyy}{Lica World, San Francisco, United States of America}

  \icmlcorrespondingauthor{Purvanshi Mehta}{purvanshi@lica.world}

  \icmlkeywords{Machine Learning, ICML}

  \vskip 0.3in
]



\printAffiliationsAndNotice{}  % no special notice (required even if empty)

\begin{abstract}

Generative video models are increasingly used in design animation tasks, yet no standardized evaluation framework exists for this domain. Unlike natural video generation, design animation imposes structured constraints: specific components shall animate with prescribed motion types, directions, speed and timing, while non-animated regions must remain stable and layout structure must be preserved. This paper provides a fully automated evaluation framework organized across four dimensions: layout fidelity, motion correctness, temporal quality, and content fidelity. This eliminates the reliance on subjective human evaluation and establishes a common basis for benchmarking progress in the field. We release the code and dataset here: \url{https://github.com/purvanshi/lica-bench}.

\end{abstract}
\section{Introduction}

Design animations are a cornerstone of modern digital communication. Producing them requires specialized expertise: a designer must select which components to animate, choose appropriate motion types (e.g., fade, slide, scale, rotate), specify direction and timing, and ensure that the overall composition remains coherent throughout. Recent advances in generative video models have opened the door to automating this workflow, with models such as Sora~\cite{sora2024} and Veo~\cite{veo_model} now being applied to design animation tasks where a static layout and per-component motion specification are provided as input. Yet, early benchmarking results reveal systematic shortcomings: models struggle to ground animation instructions to the correct visual regions, frequently apply a single dominant motion globally, animate the wrong elements, or hallucinate motion unrelated to the prompt. These failures underscore that design animation is not a subproblem of natural video synthesis but a distinct task with structured, constraint-driven requirements where specific components must animate with prescribed motion types and timing while non-animated regions remain stable and the spatial layout preserved.

\begin{figure}[t]
    \centering
    \includegraphics[width=\linewidth]{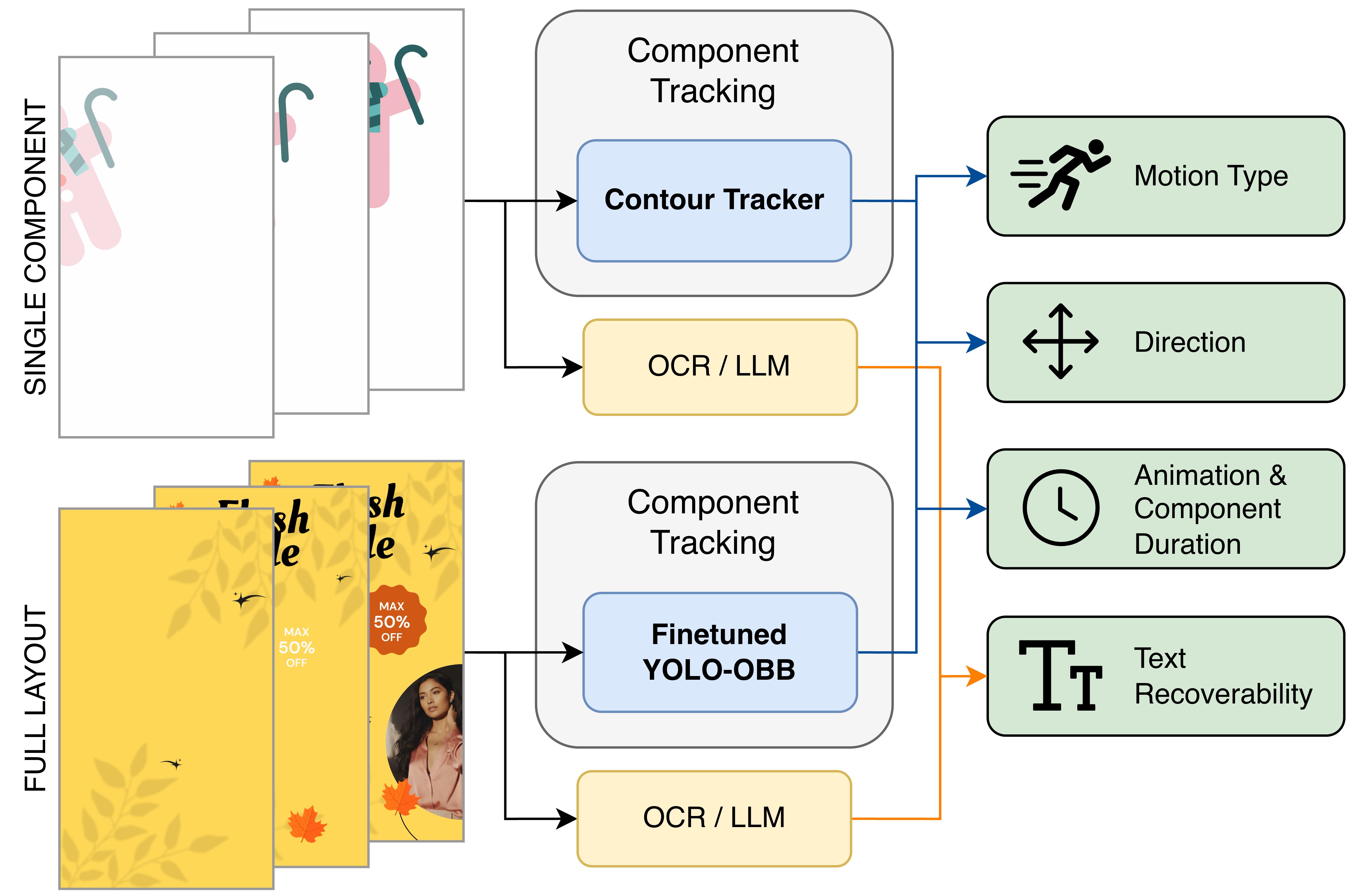}
    \caption{\textbf{Overview of the proposed framework.} We evaluate design videos across four dimensions: motion type, motion direction, duration, and text recoverability.}
    \label{fig:intro}
\end{figure}

The evaluation of generated design animations remains ad hoc: most work reports Fr\'{e}chet Video Distance or relies on small-scale human studies, neither of which captures the structured requirements unique to design animation, namely that specific components must move in prescribed ways while the rest of the composition remains intact. 

In this paper, we propose a fully automated evaluation framework organized along four complementary dimensions: layout fidelity, motion correctness, temporal quality, and content fidelity, that together provide a reproducible preliminary measure of design animation quality, eliminating the need for costly subjective assessment and establishing a common benchmark for the field.

Our key contributions are as follows.
\begin{itemize}
    \item The first dimension-decomposed evaluation framework for design animation, where each axis is independently grounded in the layout specification.
    \item Within motion correctness, a suite of four sub-metrics: motion type, direction, animation duration, and component visible duration, with ground-truth signals automatically extracted from the layout metadata.
    \item A two-track benchmark of $86$ single-component clips (isolating one element on a uniform background) and $136$ full-layout templates ($894$ component evaluations, ${\sim}6.6$ per template), spanning $29$ animation types across image, text, and group component families.
    \item Empirical validation showing that our framework recovers ground-truth motion attributes with an upper-bound partial-credit motion score of 87.8\% on single-component reference renders, with 83\% exact-match.
    \item A comparative study of Sora-2 and Veo-3.1 revealing that the two models exhibit sharply different failure modes along axes that aggregate perceptual scores cannot separate.
\end{itemize}

\section{Related Works}

\paragraph{Graphic Design Datasets and Benchmarks.}
Static graphic design research has produced a range of layout datasets, from mobile UI collections such as Rico~\cite{deka2017rico} to multi-layer template datasets like Crello~\cite{yamaguchi2021canvasvae}, poster benchmarks such as PosterLayout~\cite{hsu2023posterlayout} and CGL~\cite{zhou2022composition}, and e-commerce banner datasets~\cite{yu2024layoutdetr}. On the evaluation side, DesignProbe~\cite{lin2024designprobe} benchmarks MLLMs on color, typography, and layout understanding, while AesEval-Bench~\cite{an2026can} targets aesthetic quality assessment and Graphic-Design-Bench~\cite{deganutti2026graphic} grounds tasks in full structural metadata to enable layer-aware evaluation. Recent visual-text benchmarks further show that text rendering remains a distinct failure mode for generative models: OCRGenBench~\cite{zhang2025ocrgenbench} evaluates OCR-related generative capabilities across document, scene-text, artistic-text, and layout-rich settings, and PosterCraft~\cite{chen2025postercraft} emphasizes that high-quality poster generation requires both accurate text rendering and aesthetic layout composition. Except for~\cite{deganutti2026graphic}, these benchmarks operate on static images, leaving the temporal and motion dimensions of graphic design unaddressed.

\paragraph{Video Generation and Understanding Benchmarks.}
General video benchmarks fall into two tracks. 
VBench~\cite{huang2024vbench} decomposes generation quality into 16 hierarchical dimensions, including subject consistency, motion smoothness, and temporal flickering. validated against human preference annotations. 
EvalCrafter~\cite{liu2024evalcrafter} evaluates text-to-video models across visual, content, and motion quality dimensions using 700 prompts derived from real-world user data. 
For multimodal understanding, Video-MME~\cite{fu2025video} spans six primary visual domains with 30 subfields and varying video durations to stress-test temporal reasoning in MLLMs. 
These benchmarks target naturalistic or cinematic content and do not account for the compositional, typographic, or brand-driven properties of designed video.

\paragraph{Advertisement and Creative Video.}
A growing set of works recognizes the distinct challenges of commercial video content. 
VideoAds~\cite{zhang2025videoads} benchmarks MLLMs on advertisement videos with complex temporal structures across visual finding, summarization, and reasoning tasks, finding that even the strongest models fall well short of human performance. 
AdsQA~\cite{long2025adsqa} provides a QA benchmark derived from over 1,500 ad videos across five tasks centered on the persuasive structure of advertising. 
E-VAds~\cite{liu2026vads} targets e-commerce short videos, identifying a gap in benchmarks that address commercial intent reasoning rather than general action recognition or commonsense QA. 
While these works probe semantic and narrative understanding of ads, none evaluate the design-specific dimensions: layout, typography and motion composition, that our benchmark is built around.
\section{Dataset}
\label{sec:dataset}

Evaluating design video generation at the component level requires ground-truth data that existing video benchmarks do not provide: per-component bounding boxes, animation type and timing metadata, and deterministic reference renders against which metrics can be validated. Natural-video datasets lack layered structure entirely, and even GDB's animation evaluation set~\citep{deganutti2026graphic} comprises only 10 samples per generation task with human-only evaluation, making it insufficient for automated metric development. We therefore construct a dedicated evaluation dataset from the LICA layered-composition corpus~\citep{hirsch2026lica}, which uniquely supplies the full component-level metadata our framework requires. Specifically, we draw from the public release \texttt{lica-dataset} and filter to entries whose sub-category is \texttt{videos}, yielding 136 source layouts spanning 66 unique templates. From each layout we extract the canvas specification (dimensions, background color, total duration), and for every component its type (text, image, or group), bounding box, and animation attributes including motion type, direction, animation duration, trigger time, and visible duration. Each animated component carries exactly one motion type assigned at authoring time in LICA; we extract these assignments verbatim and do not synthesize or reassign animations. For text components we additionally extract the rendered string, font family, font size, weight, style, color, line height, letter spacing, and alignment. A component is treated as animated when it carries a non-empty \texttt{animations} list; component trees are walked recursively so that animated elements nested inside groups are not missed.

We construct evaluation data at two levels of complexity. Both tracks are rendered deterministically from the extracted layout JSON, so each sample's ground-truth video and final-frame PNG are byte-exact references against which our metrics can be validated. The \emph{full-layout} track preserves complete multi-component compositions, retaining both animated and static elements from each source layout and inheriting the full LICA motion vocabulary (up to 29 types per scene mix). All 136 video-subset layouts are kept as candidates and rendered to (layout JSON, rendered video MP4, final-frame PNG, prompt) tuples; animated component counts range from 0 to 30 per scene (mean 6.6 across all 136 layouts; 7.5 across the 120 layouts with at least one animated component). The resulting animated-component pool spans 520 image, 374 text, and 0 group elements across successful renders. The \emph{single-component} track isolates one animated element per source layout onto a white canvas to provide an atomic motion benchmark. To keep this track focused on motion types that are both visually salient and frequent enough to score reliably, we restrict it to a 15-type subset of the LICA taxonomy (\texttt{rise}, \texttt{pop}, \texttt{fade}, \texttt{wipe}, \texttt{pan}, \texttt{burst}, \texttt{ascend}, \texttt{bounce}, \texttt{tumble}, \texttt{rotate}, \texttt{drift}, \texttt{shift}, \texttt{skate}, \texttt{photorise}, \texttt{photoflow}); for each source layout we traverse all components recursively and admit only those whose pre-existing animation type falls in this allowlist, enforcing a maximum of one accepted component per layout to avoid dense duplicates from the same source scene. The retained components keep their original LICA animation parameters verbatim; we do not modify or reassign motion types, directions, or timings. To keep extraction backgrounds clean while preserving visibility, the canvas defaults to white, switching to black when a text component's color is near-white. Each accepted candidate is rendered, yielding 86 retained samples from 101 candidates, balanced across 37 image, 13 text, and 36 group components and covering 12 of the 15 allowlisted motion types (three types had no candidates clear all filters).

For each animated component, the prompt enumerates motion type,
normalized direction, animation duration, speed, trigger phase, timing
window (\texttt{from}, \texttt{duration}), and positional geometry,
plus type-specific visual fields (text content and full typographic
specification for text; alt descriptions for image and vector). Static
elements are included in the layout JSON and final-frame render but
excluded from the animated-component enumeration in the prompt. Every
generated sample preserves provenance back to its source: the manifest
records the originating template id, layout id, component ids,
component-type list, animation-type list, and all artifact paths, so
each evaluation sample is auditable to a specific LICA public dataset
entry.

\begin{table}[h]
\centering
\small
\caption{Dataset statistics. The evaluation set is drawn from the LICA videos sub-category in \texttt{lica-dataset}. The full-layout track inherits a 29-type motion vocabulary; the single-component track restricts to a 15-type subset, of which 12 are observed after candidate filtering.}
\begin{tabular}{@{}ll@{}}
\toprule
\multicolumn{2}{@{}l}{\textbf{Source data} (LICA, videos sub-category)} \\
\midrule
Video layouts / unique templates & 136 / 66 \\
Layouts with animated components & 120 \\
Animated comp.\ per layout (mean, all / animated) & 6.6 / 7.5 \\
Animated comp.\ per layout (min / max) & 0 / 30 \\
\midrule
\multicolumn{2}{@{}l}{\textbf{Full-layout track} (136 samples)} \\
\midrule
Animated components (image / text) & 520 / 374 \\
Animated comp.\ per scene (min / max / mean) & 0 / 30 / 6.6 \\
Motion types & 29 \\
\midrule
\multicolumn{2}{@{}l}{\textbf{Single-component track} (86 samples)} \\
\midrule
Component families (image / text / group) & 37 / 13 / 36 \\
Motion types (before / after filter) & 15 / 12 \\
Candidates before motion filter & 101 \\
Motion-score threshold & 0.03 \\
\bottomrule
\end{tabular}
\label{tab:dataset-stats}
\end{table}
\section{Evaluation Framework}
\label{sec:eval_framework}

The design video generation framework introduces a comprehensive evaluation framework spanning four competency dimensions: Motion Type, Direction, Duration, and Text Recoverability which reflects the natural pipeline of graphic design animation skills, enabling systematic assessment of model capabilities across a wide spectrum of practical applications.

\subsection{Tracking}

We propose two tracking frameworks, each aligned with a benchmark setting: single-component and multi-component (full) layouts. The single-component framework uses temporal differencing across frames, while the multi-component framework relies on spatial matching against known layout geometry to separate and identify multiple components on each frame independently. Both frameworks output a standardized per-frame record that contains a bounding box and confidence score for every tracked component, making the evaluation metric layer entirely tracker-agnostic.

\textbf{Contour OBB tracker (single-component layouts).}
Single-component renders sit on a white uniform background. Each frame is differenced against the rendered background color to isolate moving regions, which are then grouped into components and represented as oriented bounding boxes (OBBs). Because there is exactly one component of interest per video, no track-association step is required. This backend achieves $99.9\%$ mean detection presence across the $86$ single-component prompts for both evaluated video generation models, making it suitable as a near-ceiling reference for the metric pipeline. 

\textbf{YOLO-OBB tracker (full layouts).}
Videos in the full layout track contain multiple animated components (image and text) layered onto a background image or design. Therefore, unlike single-component layouts, the background cannot be used as a reference to track the components. To handle this, we finetune an oriented-bounding-box (OBB) variant of YOLOv11~\cite{khanam2024yolov11} on $\sim\!500\text{k}$ layouts from the LICA dataset~\cite{hirsch2026lica} with a class set of $\{\textit{IMAGE}, \textit{TEXT}\}$. Rather than linking detections across frames with a multi-object tracker, we exploit the fact that the layout metadata provides ground-truth positions for every component. For each frame, we independently run the detector and use the Hungarian algorithm to match detected OBB polygons to known layout polygons based on polygon IoU. Because the layout metadata and the generated video may differ in resolution, we apply an isotropic scaling to the layout polygons before any IoU comparison.

\subsection{Motion-Type}
\label{sec:framework_motion_type}

\textbf{Observable Motion Groups.}
The layout metadata from the LICA dataset~\cite{hirsch2026lica} contains a fine-grained vocabulary of thirty motion verbs (e.g.,~\textit{rise}, \textit{ascend}, \textit{drift}, \textit{tumble}, \textit{baseline}, \textit{burst}). Many of these verbs differ only in easing curve or timing and produce indistinguishable OBB trajectories (e.g.,~\textit{rise} and \textit{ascend} both translate the component upward). We therefore group the raw vocabulary into a smaller set of ten observable motion classes that a geometric tracker can reliably distinguish:
\[
\mathcal{T} = \left\{
\begin{aligned}
&\textit{static}, \textit{fade}, \textit{scrapbook}, \textit{pop}, \textit{wiggle}, \\
&\textit{breathe}, \textit{rotate}, \textit{pan}, \textit{sketch}, \textit{neon}
\end{aligned}
\right\}
\] 

A key design decision in this grouping concerns how the LICA dataset's reference renderer executes motion. Most animations in the LICA dataset are \emph{on-enter} effects: the component animates into its final position over a brief window (typically $0.56\text{--}1.12$\,s) and then remains static for the rest of the clip. From the tracker's perspective, this appears as a short burst of motion at the start of the clip followed by a long static plateau. This signature matches the \textit{scrapbook} class (a transient entry animation) rather than a sustained \textit{pan} (continuous translation throughout the clip). The same applies to rotational entries such as \textit{tumble}: the component rotates into place over ${\sim}1$\,s and then stops, producing a brief rotational transient rather than sustained rotation. The grouping therefore reflects the observable trajectory behavior rather than the semantic intent of the animation grammar. Animations with similar temporal signatures are consequently assigned to the same observable class:

\begin{table}[h]
\centering
\caption{Grouping of LICA motion verbs into observable motion classes.}
\label{tab:motion_groups}
\footnotesize
\setlength{\tabcolsep}{3pt}
\begin{tabular}{p{1.5cm} p{4.3cm} p{1.4cm}}
\toprule
\textbf{Behavior} & \textbf{LICA verbs} & \textbf{Class} \\
\midrule
\midrule
Translation & \textit{rise}, \textit{ascend}, \textit{drift}, \textit{pan}, \textit{wipe}, \textit{shift}, \textit{skate}, \textit{photorise}, \textit{photoflow}, \textit{scrapbook}, \textit{bounce}, \textit{stomp} & scrapbook \\ 
\midrule
Rotation & \textit{tumble}, \textit{roll}, \textit{rotate} & rotate \\
\midrule
Opacity & \textit{baseline}, \textit{fade}, \textit{flicker}, \textit{blur}, \textit{merge}, \textit{clarify}, \textit{succession}, \textit{typewriter} & fade \\
\midrule
Scale & \textit{burst}, \textit{pop} & pop \\
\midrule
Oscillation & \textit{wiggle} & wiggle \\
\midrule
Oscillation & \textit{breathe}, \textit{pulse} & breathe \\
\bottomrule
\end{tabular}
\end{table}

\begin{figure}[h]
    \centering
    \includegraphics[width=\linewidth]{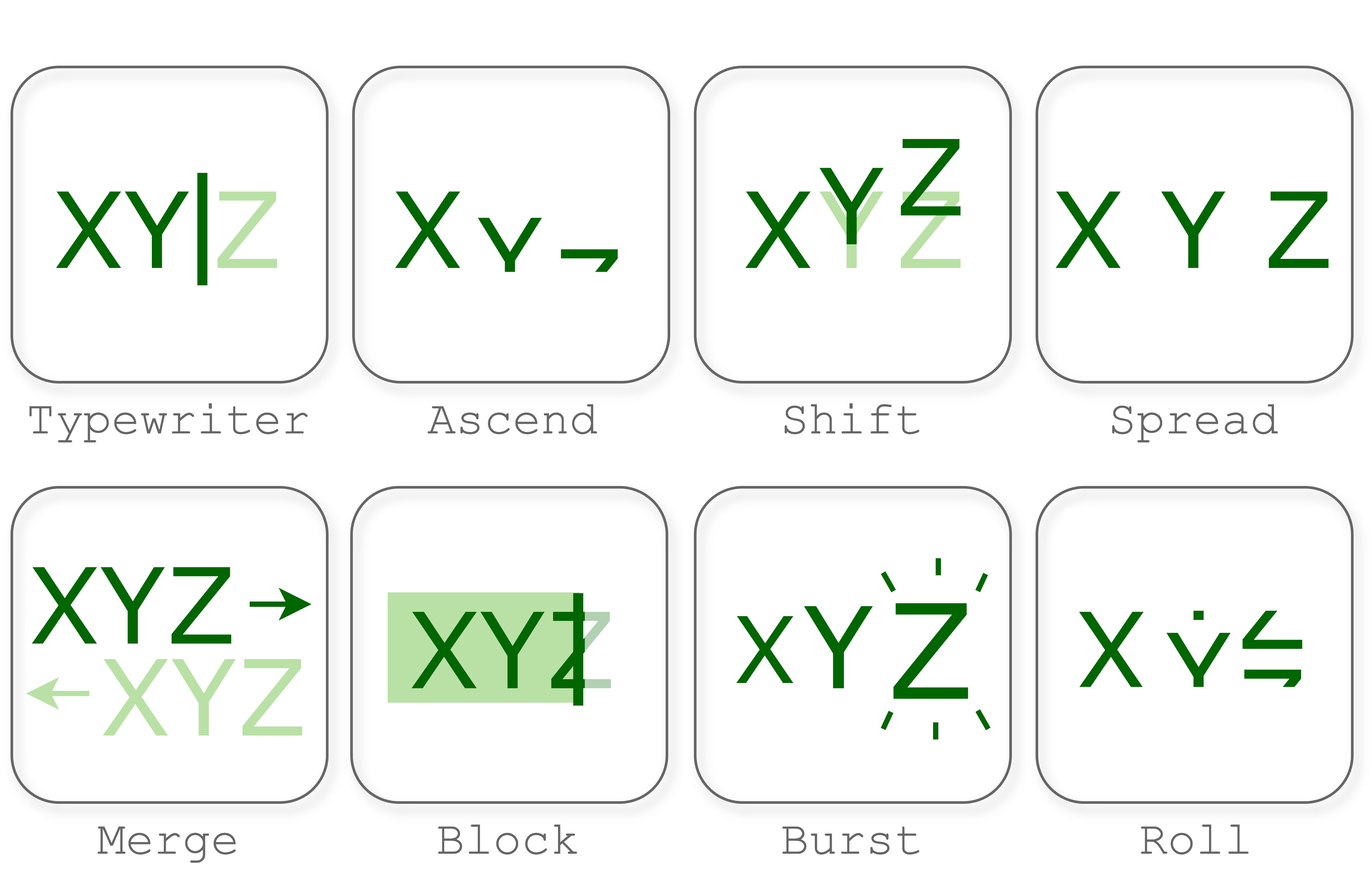}
    \caption{Visual examples of motion types.}
    \label{fig:framework_motions}
\end{figure}

\textbf{Motion-Type Classification.}
We classify the observed motion into $\hat{t} \in \mathcal{T}$ from tracked trajectory features and report accuracy with a partial-credit table for semantically related labels. We list pairs as (GT $\to$ predicted: credit): pop\,$\leftrightarrow$\,scrapbook ($0.5$), fade\,$\leftrightarrow$\,scrapbook ($0.3$), wiggle\,$\leftrightarrow$\,breathe ($0.4$), and the asymmetric entries scrapbook\,$\to$\,pan ($0.3$), rotate\,$\to$\,scrapbook ($1.0$), and scrapbook\,$\to$\,rotate ($0.5$). The asymmetric rotate$\to$scrapbook entry awards full credit because LICA's \textit{tumble} and \textit{roll} entries combine a translational entry with rotation, so a scrapbook prediction is observationally equivalent. When the ground-truth label is \textit{unknown}, the metric awards $1.0$ for predictions of \textit{static} or \textit{unknown} and $0.0$ otherwise.

The baseline classifier is a rule-based decision tree described in Table~\ref{tab:motion_rules}, grounded in design-animation terminology. The tree structure mirrors the observable motions of animation families (e.g., continuous movement for \textit{pan}, opacity changes for \textit{fade}), while the decision thresholds are chosen empirically by analyzing ground-truth examples and setting boundaries that best separate the different animation types. Energy ($E$) is computed over the start, middle, and end thirds of each clip to capture temporal dynamics (high energy = lots of motion). The noise floor of the tracker is set to $\varepsilon \approx 3 \times 10^{-3}$ such that any motion smaller than this is considered tracker jitter. All measurements are normalized by the diagonal of the canvas. An in-depth description of the motion classification rules is provided in Appendix~\ref{app:motion_rules}.

\begin{table}[t]
\centering
\caption{Overview of the priority-ordered motion classification rules. Additional details are given in Appendix~\ref{app:motion_rules}.}
\label{tab:motion_rules}
\footnotesize
\setlength{\tabcolsep}{4pt}
\begin{tabular}{cl}
\toprule
\textbf{\#} & \textbf{Rule} \\
\midrule
1  & low presence $\rightarrow$ \textit{static} \\
2  & opacity transient $\rightarrow$ \textit{fade} \\
3  & scale spike with low displacement $\rightarrow$ \textit{pop} \\
4  & entry rotation transient $\rightarrow$ \textit{rotate} \\
5  & entry motion transient $\rightarrow$ \textit{scrapbook} \\
6  & sustained rotation $\rightarrow$ \textit{rotate} \\
7  & high-frequency positional jitter $\rightarrow$ \textit{wiggle} \\
8  & periodic scale oscillation $\rightarrow$ \textit{breathe} \\
9  & sustained directional displacement $\rightarrow$ \textit{pan} \\
10 & late opacity/presence change $\rightarrow$ \textit{fade} \\
11 & residual motion energy $\rightarrow$ \textit{scrapbook} \\
12 & otherwise $\rightarrow$ \textit{static} \\
\bottomrule
\end{tabular}
\end{table}

\subsection{Motion Direction}
\label{sec:framework_direction}
Each animated component specifies a motion direction $d$ belonging to one of eight compass directions, a rotation or \textit{none} such that:
\[
d \in \left\{
\begin{gathered}
\textit{right}, \textit{left}, \textit{up}, \textit{down}, \textit{up\_right}, \textit{up\_left}, \textit{down\_right} \\
\textit{down\_left}, \textit{clockwise}, \textit{anticlockwise}, \textit{none}
\end{gathered}
\right\}
\]
We predict direction $\hat{d}$ by measuring the displacement of the component centroid and bin the movement angle to the nearest compass direction used by the layout vocabulary. The segment of the trajectory used relies on the motion-type assigned in the previous step: (a)~\textit{Transient motions} (scrapbook, pop, fade): displacement from $t=0$ to $t=N/4$, capturing the entry direction before the component settles; (b)~\textit{Pan}: net displacement over the full clip, since the component moves the entire time; (c)~\textit{Rotate}: the OBB angle series is unwrapped with period $90\degree$ (resolving the axis-swap ambiguity of \texttt{cv2.minAreaRect}) and the sign of the total angle change determines \textit{clockwise} or \textit{anticlockwise} when $|\Delta\theta| \geq 45\degree$; (d)~\textit{Non-translational motions} (static, wiggle, breathe): prediction is suppressed to \textit{none}, as these motions have no meaningful linear direction. We report accuracy by comparing the predicted direction $\hat{d}$ to the ground truth $d$.

\subsection{Animation \& Component Duration}
\label{sec:duration}

We evaluate two related but distinct timing properties. Animation duration measures how long the motion itself lasts (e.g., a component might slide in over 0.5 seconds and then remain still for the rest of the video.) Component visible duration measures how long the component is on screen, regardless of whether it is moving.

\textbf{Animation duration.}
Each layout specifies a ground-truth animation duration \(\tau_a\) (in seconds). We estimate \(\hat{\tau}_a\) from the motion-energy signal that combines per-frame changes in centroid position, scale, rotation, and opacity:
\[
    E_t = \|\Delta \mathbf{c}_t\| + 0.5\,|\Delta s_t| +
    0.3\,|\Delta\theta_t|/90 + 0.5\,|\Delta o_t|
\]
where $\Delta$ denotes the per-frame difference, $\mathbf{c}_t$ is the normalized centroid, $s_t$ the scale, $\theta_t$ the rotation, and $o_t$ the opacity proxy. After smoothing, we define the predicted duration as the time between the first and last frames where $E_t$ exceeds 15\% of its peak value. We report mean absolute error (MAE) in seconds.

\textbf{Component visible duration.} 
Each component is specified to be visible from $t_{\text{start}}$ to $t_{\text{end}}$. We estimate the visible duration $\hat{\tau}_v$ as the longest contiguous run of detector presence, converted to seconds via the frame rate. We report MAE against the ground-truth visible interval.

% ____________________________ Placed here for formatting __________________________________
\begin{table*}[t]
\centering
\caption{Motion-type classification results across both benchmark tracks and video sources. GT (\textit{italicized}) reports the evaluation ceiling on LICA renders and is excluded from best-value bolding. Single-component uses a contour OBB tracker. Full-layout uses a finetuned YOLO-OBB detector with per-frame spatial matching and is reported on the full set and tracker-reliable subset ($\text{presence} \ge 0.3$). Motion accuracy uses partial credit. MAE is in seconds. \textbf{Bold} indicates the best value per column among generators within each track block.}
\label{tab:headline}
\small
\setlength{\tabcolsep}{4pt}
\renewcommand{\arraystretch}{1.1}
\begin{tabular}{ll|cc|ccc}
\toprule
\textbf{Track} & \textbf{Model} & $n$ / Rel.{\%} & \textbf{Pres. $\uparrow$} & \textbf{Motion $\uparrow$} & \textbf{Anim. MAE $\downarrow$} & \textbf{Comp. MAE $\downarrow$} \\
\midrule
\midrule
\multirow{3}{*}{\textbf{Single Components}}
& \textit{GT}      &  \textit{86} & \textit{0.930} & \textit{0.878} & \textit{3.01} & \textit{0.65} \\
& Veo-3.1 &  86 & 0.999 & \textbf{0.649} & 4.34 & 3.99 \\
& Sora-2  &  86 & \textbf{1.000} & 0.602 & \textbf{2.41} & \textbf{3.85} \\
\midrule
\multirow{3}{*}{\textbf{Full Layouts (all)}}
& \textit{GT}      &  \textit{894} & \textit{0.759} & \textit{0.594} & \textit{2.92} & \textit{2.87} \\
& Veo-3.1 &  894 & 0.499 & \textbf{0.455} & \textbf{2.15} & \textbf{5.22} \\
& Sora-2  &  894 & \textbf{0.513} & 0.423 & 2.87 & 5.27 \\
\midrule
\multirow{3}{*}{\textbf{Full Layouts (reliable)}}
& \textit{GT}      & \textit{728 (81.4\%)} & \textit{0.921} & \textit{0.690} & \textit{3.13} & \textit{1.53} \\
& Veo-3.1 & 473 (52.9\%) & \textbf{0.892} & \textbf{0.655} & \textbf{2.69} & \textbf{3.75} \\
& Sora-2  & 499 (55.8\%) & 0.885 & 0.615 & 3.52 & 3.95 \\
\bottomrule
\end{tabular}
\end{table*}
% ___________________________________________________________________________________________

\subsection{Text Recoverability}
\label{sec:framework_text_recoverability}

For text components, content fidelity requires that the strings specified by the layout remain readable in the generated video. We evaluate this at the component level by applying OCR to sampled video frames and matching recognized text lines to the text components in \(\mathcal{L}\). We sample frames at 2 fps for text recoverability evaluation, which provides a practical trade-off between score stability and evaluation cost (Appendix~\ref{app:frame-sampling}).

Let \(\mathcal{I}_{\mathrm{text}}\) be the set of text components in \(\mathcal{L}\), with ground-truth strings \(\{g_i\}_{i \in \mathcal{I}_{\mathrm{text}}}\). For each sampled frame \(t \in \mathcal{T}(V)\), OCR returns recognized strings \(P_t=\{p_{t,1},\ldots,p_{t,K_t}\}\). We use character error rate as the matching cost,
\[
d_{\mathrm{CER}}(g,p)=
\frac{\mathrm{Lev}(g,p)}{\max(|g|,1)}.
\]
Here, \(\mathrm{Lev}(g,p)\) denotes the Levenshtein distance between strings \(g\) and \(p\), and \(|g|\) is the number of characters in the ground-truth string.

For each frame, we build a cost matrix between layout strings and OCR lines and solve a minimum-cost bipartite assignment:
\[
\pi_t^*
=
\arg\min_{\pi_t}
\sum_{i \in \mathcal{I}_{\mathrm{text}}}
d_{\mathrm{CER}}(g_i,p_{t,\pi_t(i)}).
\]
When fewer OCR lines than layout strings are detected, the OCR side is padded with dummy columns of cost \(1\). Extra OCR lines that are not assigned to any layout string are ignored. This frame-wise assignment handles multiple text components without relying on OCR reading order.

After frame-wise assignment, each layout text component selects its best assigned OCR match over time:
\[
\begin{aligned}
A_{\mathrm{text}}(V,\mathcal{L})
&=
\frac{1}{|\mathcal{I}_{\mathrm{text}}|}
\sum_{i \in \mathcal{I}_{\mathrm{text}}}
\max_{t \in \mathcal{T}(V)} \\
&\quad \left[1 - d_{\mathrm{CER}}(g_i, p_{t,\pi_t^*(i)})\right]_+ .
\end{aligned}
\]
This aggregation allows different text components to become readable in different frames. We report this score as Best AR (best accuracy rate), since it averages the best assigned recognition score for each text component over time. We also report Hard AR, \(H_{\mathrm{text}}\), defined as the fraction of layout text components whose assigned candidate is an exact match in at least one sampled frame.
\section{Experiments}
\label{sec:experiments}

We evaluate two state-of-the-art video generation models, Sora-2~\cite{sora2024} and Veo-3.1~\cite{veo_model}, on both benchmark tracks (single-component and full-layout). We additionally run the full pipeline on ground-truth videos from the LICA dataset~\cite{hirsch2026lica} to establish an upper bound on what the tracker/classifier cascade can achieve, separating metric-pipeline limitations from generator limitations.

\subsection{Experimental Setup}
Both models receive identical layout specifications and prompt formulations; the resulting clips are scored without any per-model tuning of thresholds in the evaluation pipeline. The single-component track contains $86$ videos per generator. The full-layout track contains $136$ videos per generator, yielding $894$ component-level evaluations per generator, averaging $7.5$ animated components per layout. Single-component videos use the contour OBB tracker (Section~\ref{sec:eval_framework}), which is deterministic and model-free. Full-layout videos use the finetuned YOLO-OBB tracker as the primary backend. We define a component evaluation as \emph{tracker-reliable} when its detection presence fraction exceeds $0.3$ (i.e., the component is detected in at least $30\%$ of frames). We report metrics in Table~\ref{tab:headline} on both the \emph{all-components} pool and the \emph{tracker-reliable} subset and treat the gap between them as a first-class evaluation signal.

\subsection{Evaluation Pipeline Validation}

Before comparing generated videos, we validate the evaluation pipeline on LICA reference renders. For motion type, direction, duration, and text recoverability, the GT entries in Tables~\ref{tab:headline},~\ref{tab:direction},~\ref{tab:duration} and~\ref{tab:generator-text-benchmark} estimate the ceiling imposed by the tracker/classifier pipeline when the video exactly follows the layout specification.

\subsection{Motion-Type Classification.} \label{exp:motion-type-classification}
The results on motion-type classification are given in Table~\ref{tab:headline}. We provide confusion matrices for all models in each track in Appendix~\ref{app:confusion-matrices}. We break down the sources of error below.

\textbf{Single-component.} 83\% of samples receive a perfect score. The remaining 12\% of errors come from three confusable cases: (i)~\textit{fade}$\to$\textit{scrapbook}, short fade-ins induce transient centroid jitter from changing contours; (ii)~\textit{pop}$\leftrightarrow$\textit{scrapbook}, objects simultaneously scale up and move into place, making it difficult to distinguish scaling motion from entry translation; and (iii)~three low-amplitude cases collapsing to \textit{static} below the tracker noise floor. These errors that persist on ground-truth renders reflect the limitation of the tracker signal (e.g., opacity is only indirectly observed via bbox changes).

\textbf{Full-layout.} The 19-point drop from single-component ($0.878$) to full-layout reliable ($0.690$) is dominated by the collapse of fade- and pop-family components onto the \textit{scrapbook} class. Per-class GT accuracy is $0.957$ for \textit{scrapbook}, $0.946$ for \textit{rotate}, $0.489$ for \textit{pop}, $0.373$ for \textit{fade}, $0.154$ for \textit{breathe}, and $0.0$ for \textit{static}. Of the $340$ non-perfect predictions on GT full-layout, three confusions account for $83\%$ of them: (i)~\textit{fade}$\to$\textit{scrapbook} ($156$, $46\%$ of non-perfect), (ii)~\textit{pop}$\to$\textit{scrapbook} ($127$, $37\%$), and (iii)~\textit{rotate}$\to$\textit{scrapbook} ($67$, full credit but visually equivalent). All three failures stem from missing motion cues: the YOLO-OBB tracker has no opacity signal, so fades appear as small bbox jitter and are classified as scrapbook; multi-component pops weaken the scale-pulse signature; and tumble/roll entry motions look indistinguishable from translational scrapbook entries.

\textbf{Generator accuracy.} Sora-2 reaches $0.615$ and Veo-3.1 reaches $0.655$ on the tracker-reliable subset of full-layout, both close to the GT ceiling of $0.690$. The remaining gap is explained by two additive effects: (i)~the video generator under-animates the prescribed motion (producing weaker signals that more often fall below thresholds), and (ii)~the generator adds spurious motion to supposedly-static components, introducing false positives that the GT renders do not exhibit.

\paragraph{Tracker backend effect.} We compare the finetuned YOLO-OBB against a pretrained YOLOv11x-OBB (DOTA-v1 weights, no domain-specific training) on the full-layout track. Because the matcher is purely spatial polygon IoU and does not gate on detection class, the pretrained model is mechanically able to hand detections to the layout-component matcher; in practice, however, design-layout text and graphical components do not look like the aerial-image classes the pretrained model was trained on, so its detections are sparse and intermittent. Concretely, the finetuned tracker recovers $2.7\text{--}2.9{\times}$ more reliable components than the pretrained on GT ($728$ vs.\ $266$) and Sora-2 ($499$ vs.\ $169$), and $1.7{\times}$ more on Veo-3.1 ($473$ vs.\ $277$); it also yields higher per-component presence on the reliable subset ($0.921$ vs.\ $0.782$ on GT). Both effects feed directly into motion-type accuracy on the tracker-reliable subset, where the finetuned backend leads by $8.3$ points on GT ($0.690$ vs.\ $0.607$), $16.2$ points on Sora-2 ($0.615$ vs.\ $0.453$), and $10.0$ points on Veo-3.1 ($0.655$ vs.\ $0.555$). The pretrained run is reported in this paragraph as a baseline for whether domain-specific tracker training was necessary at all.

\subsection{Motion Direction}

We report direction accuracy in Table~\ref{tab:direction}.

\begin{table}[h]
\centering
\caption{Motion direction evaluation on single-component and full-layout tracks. \emph{Motion detected} is the fraction of directional components for which the classifier correctly predicts any compass label other than \textit{none}. \emph{Correctly none} is the fraction of non-directional components where the classifier correctly predicts \textit{none}. \emph{Correct half} is the fraction of pan-family components whose predicted compass label falls in the correct horizontal half (left-family or right-family) relative to the LICA-canonical direction. Full-layout results use only tracker-reliable components, hence denominators vary by model.
}
\label{tab:direction}
\resizebox{\columnwidth}{!}{
\begin{tabular}{lccc}
\toprule
\textbf{Metric} & \textbf{GT} & \textbf{Sora-2} & \textbf{Veo-3.1} \\
\midrule
\multicolumn{4}{l}{\textit{Single Component}} \\
\midrule
Motion detected & \textit{54/58 (93.1\%)} & 40/58 (69.0\%) & \textbf{42/58 (72.4\%)} \\
Correctly `none' & \textit{18/28 (64.3\%)} & \textbf{10/28 (35.7\%)} & 4/28 (14.3\%) \\
Correct half (pan) & \textit{12/40 (30.0\%)} & \textbf{14/40 (35.0\%)} & \textbf{14/40 (35.0\%)} \\
Rotate CW/CCW & \textit{1/18 (5.6\%)} & 0/18 (0.0\%) & 0/18 (0.0\%) \\
Exact acc. & \textit{26/86 (30.2\%)} & \textbf{19/86 (22.1\%)} & 8/86 (9.3\%) \\
\midrule
\multicolumn{4}{l}{\textit{Full Layout (all)}} \\
\midrule
Motion detected & \textit{318/416 (76.4\%)} & 200/416 (48.1\%) & \textbf{235/416 (56.5\%)} \\
Correctly `none' & \textit{116/478 (24.3\%)} & \textbf{264/478 (55.2\%)} & 212/478 (44.4\%) \\
Correct half (pan) & \textit{98/316 (31.0\%)} & 46/316 (14.6\%) & \textbf{75/316 (23.7\%)} \\
Rotate CW/CCW & \textit{1/100 (1.0\%)} & 0/100 (0.0\%) & 0/100 (0.0\%) \\
Exact acc. & \textit{164/894 (18.3\%)} & \textbf{287/894 (32.1\%)} & 256/894 (28.6\%) \\
\midrule
\multicolumn{4}{l}{\textit{Full Layout (tracker-reliable (presence $\geq$ 0.3))}} \\
\midrule
Motion detected & \textit{303/332 (91.3\%)} & 157/207 (75.8\%) & \textbf{177/203 (87.2\%)} \\
Correctly `none' & \textit{59/396 (14.9\%)} & \textbf{120/292 (41.1\%)} & 53/270 (19.6\%) \\
Correct half (pan) & \textit{92/258 (35.7\%)} & 39/162 (24.1\%) & \textbf{60/157 (38.2\%)} \\
Rotate CW/CCW & \textit{1/74 (1.4\%)} & 0/45 (0.0\%) & 0/46 (0.0\%) \\
Exact acc. & \textit{105/728 (14.4\%)} & \textbf{139/499 (27.9\%)} & 85/473 (18.0\%) \\
\bottomrule
\end{tabular}
}
\end{table}

The direction classifier reliably detects the presence of directional motion: on GT single-components it identifies non-zero displacement in $93.1\%$ of directional components, and on the tracker-reliable subset of GT full-layouts it still recovers detectable directional motion in $91.3\%$ of cases. For pan-family components, the ``correct half'' metric is well above the $25\%$ chance baseline of an 8-way bin on both tracks ($30\%$ single-component, $35.7\%$ full-layout) but is limited by the short LICA entry transient: the ${\sim}17$ frames of motion are often dominated by vertical ``rise-into-place'' dynamics, which overwhelm the weaker horizontal signal controlled by the direction parameter. Rotate CW/CCW accuracy is very low, reflecting a fundamental limitation of OBB-based angle extraction that encodes axis orientation rather than rotation direction; the converse failure mode is also visible on the full-layout reliable pool ($728$ components), where the GT ``correctly none'' rate is only $14.9\%$ because multi-component scenes accumulate enough incidental drift to cross the $1\%$-diagonal threshold on non-directional components. Despite these limitations, the method succeeds at its primary design goal as it provides a per-component, automated indicator of whether the generator produced directional motion at all, a signal that frame-level perceptual metrics cannot surface.

Generator outputs approach the GT ceiling on the full-layout track (Sora-2 $75.8\%$, Veo-3.1 $87.2\%$ vs.\ $91.3\%$ GT), but remain noticeably below it on single-component motion ($69$--$72\%$ vs.\ $93\%$ GT). Veo-3.1 slightly exceeds the GT ceiling on pan-direction fidelity ($38.2\%$ vs.\ $35.7\%$ GT), while Sora-2 achieves the highest exact accuracy ($27.9\%$) mainly because its weaker entry transients are more often classified as \textit{none}.

\subsection{Duration Metrics.}

We report two duration metrics: \emph{animation duration} (how long the component exhibits active motion) and \emph{component visible duration} (how long the component is on-screen). The energy-based animation-duration estimator (Section~\ref{sec:duration}) measures the temporal extent of above-noise motion energy; the component-duration estimator uses the first-to-last detection window. Table~\ref{tab:duration} reports MAE, median absolute error, and signed bias for both metrics.

\begin{table}[h]
\centering
\caption{Duration estimation errors (seconds). Single-component results use all $86$ samples; full-layout results are restricted to the tracker-reliable subset.}
\label{tab:duration}
\resizebox{\columnwidth}{!}{
\begin{tabular}{l|ccc|ccc}
\toprule
\multirow{2}{*}{\textbf{Model}} & \multicolumn{3}{c}{\textbf{Animation Duration}} & \multicolumn{3}{c}{\textbf{Component Duration}} \\
\cmidrule(lr){2-4}\cmidrule(lr){5-7}
 & MAE \(\downarrow\) & Med.\ AE \(\downarrow\) & Bias & MAE \(\downarrow\) & Med.\ AE \(\downarrow\) & Bias \\
\midrule
\midrule
\multicolumn{7}{l}{\textit{Single Component}} \\
\midrule
  \textit{GT}      & \textit{3.01} & \textit{0.53} & \textit{$+$2.61} & \textit{0.65} & \textit{0.22} & \textit{$-$0.65} \\
  Sora-2  & \textbf{2.41} & \textbf{2.94} & \textbf{$+$2.11} & \textbf{3.85} & \textbf{1.76} & $-$3.45 \\
  Veo-3.1 & 4.34 & 5.36 & $+$4.27 & 3.99 & 3.00 & \textbf{$+$0.49} \\
\midrule
\multicolumn{7}{l}{\textit{Full Layout (all)}} \\
\midrule
 \textit{GT}    & \textit{2.92} & \textit{1.64} & \textit{$+$2.65} & \textit{2.87} & \textit{0.89} & \textit{$-$1.69} \\
 Sora-2  & 2.87 & 2.34 & $+$2.50 & 5.27 & \textbf{4.29} & \textbf{$-$3.93} \\
 Veo-3.1 & \textbf{2.15} & \textbf{1.22} & \textbf{$+$1.62} & \textbf{5.22} & 4.54 & $-$3.95 \\
\midrule
\multicolumn{7}{l}{\textit{Full Layout (reliable)}} \\
\midrule
  \textit{GT}      & \textit{3.13} & \textit{2.20} & \textit{$+$2.94} & \textit{1.53} & \textit{0.57} & \textit{$-$0.55} \\
  Sora-2  & 3.52 & 2.96 & $+$3.35 & 3.95 & 1.51 & \textbf{$-$1.56} \\
  Veo-3.1 & \textbf{2.69} & \textbf{2.31} & \textbf{$+$2.46} & \textbf{3.75} & \textbf{1.41} & $-$2.09 \\
\bottomrule
\end{tabular}
}
\end{table}

On single-component GT renders, component-duration MAE is only $0.65$\,s thanks to a clean contour-tracker lock; animation duration is harder (median AE $0.53$\,s, mean inflated by a heavy tail) because post-animation contour jitter and settling drift continue to register above the energy threshold. On the full-layout track, component MAE rises to $1.53$\,s because the YOLO tracker occasionally loses and re-acquires components, fragmenting the detection window; animation MAE is $3.13$\,s (bias $+2.94$\,s), consistent with the same over-estimation mechanism.

The single-component and full-layout tracks expose opposite ordering between Sora-2 and Veo-3.1 on animation duration. Sora-2 performs better on single-component clips ($2.41$,s vs.\ $4.34$,s MAE) because its systematically shorter outputs truncate the motion envelope, while on full-layout Veo-3.1 achieves the lowest duration error ($2.69$,s vs.\ $3.52$,s) as Sora-2’s weaker motions more often fall below the energy threshold.

\subsection{Text Recoverability}

Text recoverability has an additional source of evaluator variation because candidate strings can be extracted either by OCR or by an LLM-based transcription model. We therefore validate both text evaluators on reference-rendered scenes, where the target text is known to be present (Table~\ref{tab:generator-text-benchmark}). In both cases, the evaluator only extracts candidate strings; the final score is computed by the same CER-based assignment metric defined in Section~\ref{sec:framework_text_recoverability}. Unlike the motion and layout metrics above, text recoverability is computed directly from video frames sampled at 2 FPS, following Appendix~\ref{app:frame-sampling}, and does not require tracker-reliable component detections. We evaluate text content on paired Sora-2 and Veo-3.1 generations, reporting results separately for single-component and full-layout scenes and under both OCR and LLM evaluators.

\begin{table}[h]
\caption{Text recoverability of frontier video generation models on design-centric text rendering tasks. We compare Sora-2 and Veo-3.1 across controlled single-component scenes and full-layout LICA video scenes, using both evaluator types described in Section~\ref{sec:framework_text_recoverability}.}
\centering
\small
\begin{tabular}{l|l|ccc}
\toprule
Dataset & Model & Evaluator & Best AR \(\uparrow\) & Hard AR \(\uparrow\) \\
\midrule
\midrule
\multirow{6}{*}{\shortstack{\textit{Single}\\\textit{Compnt.}}} & \multirow{2}{*}{\textit{GT}} & \textit{OCR} & \textit{0.769} & \textit{0.769} \\
 & & \textit{LLM} & \textit{1.000} & \textit{1.000} \\
\cmidrule(lr){2-5}
& \multirow{2}{*}{Sora-2} & OCR & \textbf{0.941} & \textbf{0.769} \\
& & LLM & \textbf{0.977} & \textbf{0.846} \\
\cmidrule(lr){2-5}
 & \multirow{2}{*}{Veo-3.1} & OCR & 0.907 & 0.692 \\
& & LLM & 0.898 & 0.615 \\
\midrule
\multirow{6}{*}{\shortstack{\textit{Full}\\\textit{Layout}\\\textit{(all)}}} & \multirow{2}{*}{\textit{GT}} & \textit{OCR} & \textit{0.862} & \textit{0.659} \\
 & & \textit{LLM} & \textit{0.960} & \textit{0.940} \\
\cmidrule(lr){2-5}
 & \multirow{2}{*}{Sora-2} & OCR & \textbf{0.817} & \textbf{0.586} \\
 & & LLM & \textbf{0.859} & \textbf{0.671} \\
\cmidrule(lr){2-5}
 & \multirow{2}{*}{Veo-3.1} & OCR & 0.659 & 0.389 \\
 & & LLM & 0.669 & 0.416 \\
\bottomrule
\end{tabular}
\label{tab:generator-text-benchmark}
\end{table}

Text recoverability reveals a clear gap between reference renders and generated videos, especially in full-layout scenes. Although the LLM evaluator reaches $0.940$ Hard AR on full-layout reference renders, the best generated result reaches only $0.671$ (Sora-2 with LLM), indicating that current generators often fail to preserve required text exactly in multi-component compositions. Sora-2 consistently outperforms Veo-3.1 on the full-layout track under both evaluators, while the smaller and backend-dependent gap on single-component prompts suggests that isolated text rendering is less diagnostic than text preservation in complete layouts.

\section{Conclusion}

We have presented a fully automated, dimension-decomposed evaluation framework for generative design animation that measures motion type, direction, duration, and layout fidelity against ground-truth layout specifications without human judgment. Applied to Sora-2 and Veo-3.1, the framework exposes per-component failure modes such as under-animation, directional inaccuracy, and duration bias, that aggregate perceptual metrics conceal. We note that several of these axes remain limited even on ground-truth reference renders, particularly full-layout motion classification, direction accuracy, and duration estimation. Therefore we position these as diagnostic signals that provide a preliminary evaluation of design animation. Future work will extend the motion-correctness axis to character-level evaluation, and will incorporate learned detectors to complement the current rule-based classifier on motion families where geometric features alone are insufficient.

\section{Impact Statement}
The contribution of this paper is measurement rather than a new capability or generative artifact. By quantifying where frontier generative and vision-language models succeed and fail on design animation tasks, it is intended to help researchers, designers, and tool-builders make better-informed decisions about when these models are, and are not, ready to act as collaborators. Reliable design animation competence is also a prerequisite for accessibility-sensitive deployments, so exposing current gaps supports responsible use in professional workflows. We view this as supportive of human-AI co-creativity in its augmentation sense: surfacing current limitations so that design tools are deployed as aids to designers rather than as replacements. Beyond the failure modes reported in our experiments, we do not foresee specific additional societal consequences that warrant highlighting here.

\bibliography{references}
\bibliographystyle{icml2026}

\newpage
\appendix
\onecolumn

%================================================================
\section{Dataset Examples}
\label{app:dataset-examples}

We provide representative samples from both evaluation tracks. Three from the full layout
track (Section~\ref{app:full-examples}) and eight from the
single-component track (Section~\ref{app:single-examples}). To keep the appendix compact, we
provide the verbatim Track~1 prompt for one full-layout example
(Example~1, Section~\ref{app:full-1}) and one single-component example
(S3, Section~\ref{app:single-examples}); all other prompts in our dataset
follow the same schema and differ only in their per-component metadata.

%================================================================
\subsection{Full-Layout Examples}
\label{app:full-examples}

Each full-layout example is presented as a frame-by-frame animation sampled at the
entrance windows of its components, with a summary table reporting canvas
size, total component count, animated component count, and the set of
animation primitives invoked. The verbatim Track~1 prompt is given for
Example~1 to illustrate the prompt format; Examples~2 and~3 use the same
schema. Static elements (background, decorative vectors) are present in the
rendered video and visible in the final frame but, following the convention
introduced in Section~3, are excluded from the animated-component
enumeration in the prompt.

%--------------------- FULL LAYOUT 1 ----------------------------
\subsubsection{Full Layout 01}
\label{app:full-1}

\begin{figure}[h!]
\centering
\includegraphics[width=0.95\textwidth]{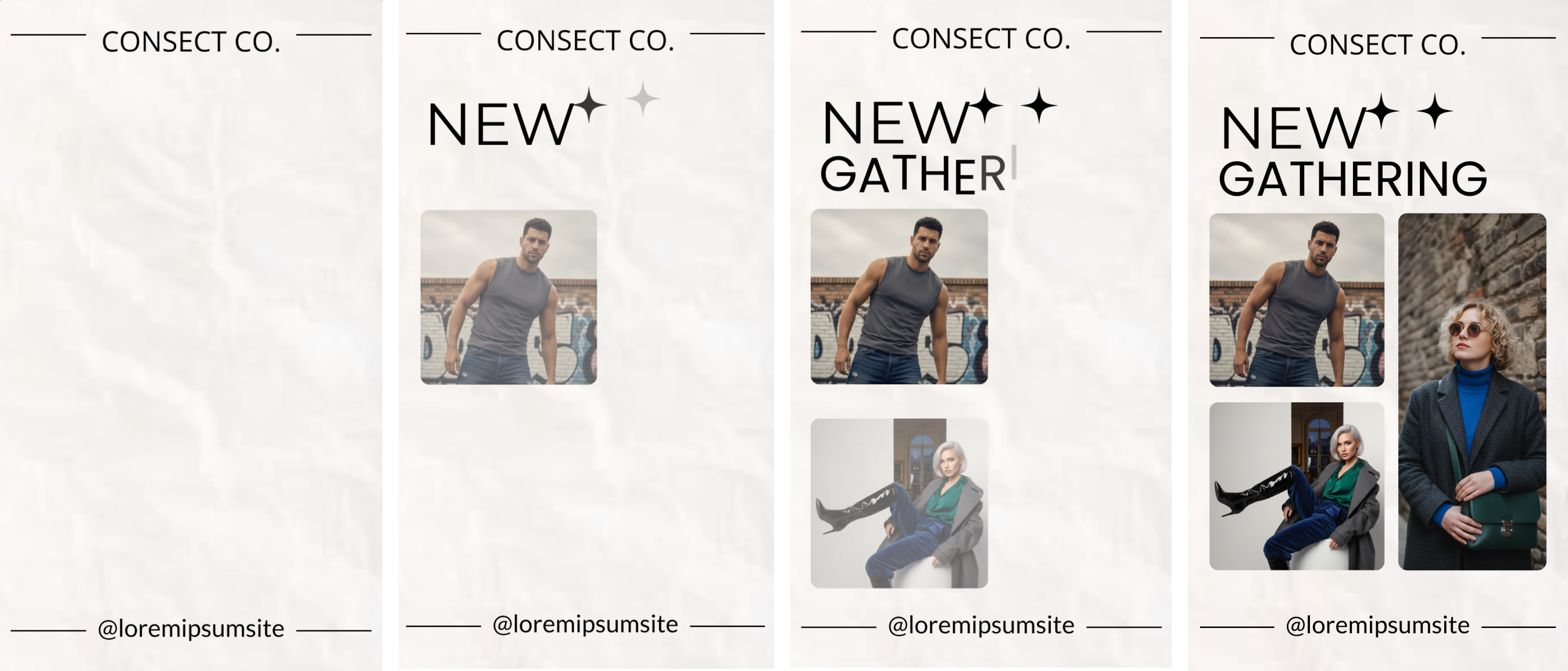}
\caption{Animation of \texttt{Full Layout 01}.
The four entrance cohorts span a 37.1\,s clip: opening cohort at $t{=}0$\,s
(thumbnails \textsc{0-5}, \textsc{0-6} via \texttt{rise down}; text
\textsc{0-9} via \texttt{ascend up}; text \textsc{0-10} via \texttt{bounce}),
followed by group~\textsc{0-11} (\texttt{pan right}) at $t{=}13.13$\,s,
group~\textsc{0-12} (\texttt{rise up}) at $t{=}24.38$\,s, and
group~\textsc{0-13} (\texttt{pan left}) at $t{=}34.38$\,s.}
\label{fig:full-1-storyboard}
\end{figure}

\begin{table}[h!]
\centering
\small
\begin{tabular}{ll}
\toprule
\textbf{Property} & \textbf{Value} \\
\midrule
Canvas             & $1080{\times}1920$, background \texttt{rgb(252,246,243)} \\
Total components (recursive) & 18 \\
Animated components & 7 (2~\textsc{image}, 2~\textsc{text}, 3~\textsc{group}) \\
Animation primitives & \texttt{rise}, \texttt{ascend}, \texttt{bounce}, \texttt{pan} \\
Directions invoked   & \texttt{down}, \texttt{up}, \texttt{right}, \texttt{left} \\
Entrance regime    & Staggered ($t{=}0,\,13.13,\,24.38,\,34.38$\,s) \\
\bottomrule
\end{tabular}
\caption{Summary of full-layout Example~1.}
\label{tab:full-1-summary}
\end{table}

\paragraph{Track~1 prompt (verbatim).}
 
{\scriptsize
\begin{Verbatim}[breaklines=true,breakanywhere=true]
Generate an animated video at 1080x1920 pixels with background color rgb(252, 246, 243).

Components (14 total) with the following exact specifications:
- Component 0-0-0 (IMAGE):
  - Position: left=-87.6px, top=0.0px (top-left corner of bounding box)
  - Size: width=1255.2px, height=1920.0px
  - Element attributes: description=This image features an abstract texture, dominated by shades of white and very light grey. It depicts a soft, crumpled surface, possibly paper or fabric, with numerous gentle folds, creases, and subtle shadows that create an undulating, organic pattern. The style is minimalist and textural, emphasizing the subtle variations in light and shadow across the uneven surface., has source asset
  - Animation: none (static component)
  - The component remains visible from t=0s to t=3.7361380000000004s

- Component 0-1 (IMAGE):
  - Position: left=834.019px, top=111.918px (top-left corner of bounding box)
  - Size: width=210.939px, height=4px
  - Element attributes: description=This image appears to be entirely transparent, showing no visible content, subjects, or colors. It is essentially a blank canvas, with no discernible patterns, shapes, or text present within its frame., has source asset
  - Animation: none (static component)
  - The component remains visible from t=0s to t=3.7361380000000004s

- Component 0-2 (IMAGE):
  - Position: left=834.019px, top=1784.13px (top-left corner of bounding box)
  - Size: width=210.939px, height=4px
  - Element attributes: description=The image is entirely transparent, revealing no discernible content or visual information. There are no main subjects, colors, patterns, text, or any other visible elements to describe within the frame. It presents as a completely empty and featureless space., has source asset
  - Animation: none (static component)
  - The component remains visible from t=0s to t=3.7361380000000004s

- Component 0-3 (IMAGE):
  - Position: left=35.042px, top=111.918px (top-left corner of bounding box)
  - Size: width=210.939px, height=4px
  - Element attributes: description=The image is completely black, which indicates a transparent or empty space. There are no visible subjects, colors, or discernible elements within the frame. No text or specific style can be identified., has source asset
  - Animation: none (static component)
  - The component remains visible from t=0s to t=3.7361380000000004s

- Component 0-4 (IMAGE):
  - Position: left=35.042px, top=1784.13px (top-left corner of bounding box)
  - Size: width=210.939px, height=4px
  - Element attributes: description=The image provided is entirely black, appearing transparent. There are no discernible subjects, colors, patterns, or text visible within it. It conveys no visual information., has source asset
  - Animation: none (static component)
  - The component remains visible from t=0s to t=3.7361380000000004s

- Component 0-5 (IMAGE):
  - Position: left=497.359px, top=269.552px (top-left corner of bounding box)
  - Size: width=104.189px, height=104.189px
  - Element attributes: description=The image provided is completely transparent, resembling an empty space or a blank canvas. There are no visible subjects, colors, text, or any discernible elements whatsoever. It appears as an absence of visual information, offering no content to describe beyond its transparency., has source asset
  - Animation:
    - Motion type: rise
    - Direction: down
    - Animation duration: 0.56s
    - Speed: 1
    - Trigger: onEnter (begins at t=0s)
    - The component remains visible until t=3.89s

- Component 0-6 (IMAGE):
  - Position: left=649.59px, top=269.552px (top-left corner of bounding box)
  - Size: width=104.189px, height=104.189px
  - Element attributes: description=The image is completely black, presenting a solid and uniform absence of light. No discernible subjects, colors, patterns, or textual elements are visible within this dark frame, appearing as a blank space., has source asset
  - Animation:
    - Motion type: rise
    - Direction: down
    - Animation duration: 0.56s
    - Speed: 1
    - Trigger: onEnter (begins at t=0s)
    - The component remains visible until t=3.89s

- Component 0-7 (TEXT):
  - Content: "CONSECT CO."
  - Position: left=274.804px, top=76.2693px (top-left corner of bounding box)
  - Size: width=530.392px, height=95.1365px
  - Font: Open Sans--400, size 79px, weight 400, style normal
  - Color: rgb(1, 1, 1)
  - Line height: 111.0px, letter spacing: 0em, text alignment: center
  - Animation: none (static component)
  - The component remains visible from t=0s to t=3.7361380000000004s

- Component 0-8 (TEXT):
  - Content: "@loremipsumsite"
  - Position: left=194.528px, top=1728.92px (top-left corner of bounding box)
  - Size: width=690.944px, height=83.0784px
  - Font: Lato--400, size 69px, weight 400, style normal
  - Color: rgb(1, 1, 1)
  - Line height: 97.0px, letter spacing: 0em, text alignment: center
  - Animation: none (static component)
  - The component remains visible from t=0s to t=3.7361380000000004s

- Component 0-9 (TEXT):
  - Content: "NEW "
  - Position: left=85.1248px, top=252.865px (top-left corner of bounding box)
  - Size: width=464.329px, height=199.123px
  - Font: Montserrat--400, size 166px, weight 400, style normal
  - Color: rgb(1, 1, 1)
  - Line height: 232.4px, letter spacing: 0em, text alignment: left
  - Animation:
    - Motion type: ascend
    - Direction: up
    - Animation duration: 0.7s
    - Speed: 1
    - Trigger: onEnter (begins at t=0s)
    - The component remains visible until t=3.89s

- Component 0-10 (TEXT):
  - Content: "GATHERING"
  - Position: left=85.0673px, top=417.067px (top-left corner of bounding box)
  - Size: width=913.974px, height=167.742px
  - Font: Poppins--400, size 139px, weight 400, style normal
  - Color: rgb(1, 1, 1)
  - Line height: 195.7px, letter spacing: 0em, text alignment: left
  - Animation:
    - Motion type: bounce
    - Direction: unknown
    - Animation duration: 1.55s
    - Speed: unknown
    - Trigger: onEnter (begins at t=0s)
    - The component remains visible until t=3.89s

- Component 0-11-0 (IMAGE):
  - Position: left=61.6317px, top=611.289px (top-left corner of bounding box)
  - Size: width=497.341px, height=493.2339999999999px
  - Element attributes: description=A young woman with long blonde hair partially covering her face strikes a dynamic pose, dressed in a stylish white high-neck crop top and matching bottom. The image has a warm, slightly desaturated filter, giving it a vintage or artistic aesthetic. She is set against a light-colored wall with prominent dark, parallel vertical lines, adding a modern architectural element to the background., has source asset
  - Animation:
    - Motion type: pan
    - Direction: right
    - Animation duration: 0.56s
    - Speed: 1
    - Trigger: onEnter (begins at t=13.126650000000001s)
    - The component remains visible until t=16.57665s

- Component 0-12-0 (IMAGE):
  - Position: left=61.6317px, top=1147.52px (top-left corner of bounding box)
  - Size: width=497.341px, height=477.308px
  - Element attributes: description=A stylish woman is seated against a plain, light gray background, striking a chic pose with one leg raised, highlighting her footwear. She wears a black and white animal print top paired with loose-fitting beige trousers and knee-high tan heeled boots. A fluffy, cream-colored jacket is draped over her arm, completing her sophisticated and fashionable ensemble., has source asset
  - Animation:
    - Motion type: rise
    - Direction: up
    - Animation duration: 0.56s
    - Speed: 1
    - Trigger: onEnter (begins at t=24.37881s)
    - The component remains visible until t=27.44881s

- Component 0-13-0 (IMAGE):
  - Position: left=597.686px, top=611.289px (top-left corner of bounding box)
  - Size: width=420.682px, height=1013.54px
  - Element attributes: description=A stylish woman stands against a muted orange wall, looking slightly upwards with a composed expression. She wears a chic coat with large geometric patterns in shades of beige, black, and dark brown, paired with a thick black scarf and trendy dark-rimmed sunglasses. She holds a structured, light tan shoulder bag in front of her, completing her elegant and autumnal ensemble., has source asset
  - Animation:
    - Motion type: pan
    - Direction: left
    - Animation duration: 0.56s
    - Speed: 1
    - Trigger: onEnter (begins at t=34.38075s)
    - The component remains visible until t=37.12075s

The rest of the canvas must remain a solid rgb(252, 246, 243) background with no other elements.

\end{Verbatim}
}

%--------------------- FULL LAYOUT 2 ----------------------------
\subsubsection{Full Layout 02}
\label{app:full-2}

\begin{figure}[h!]
\centering
\includegraphics[width=0.95\textwidth]{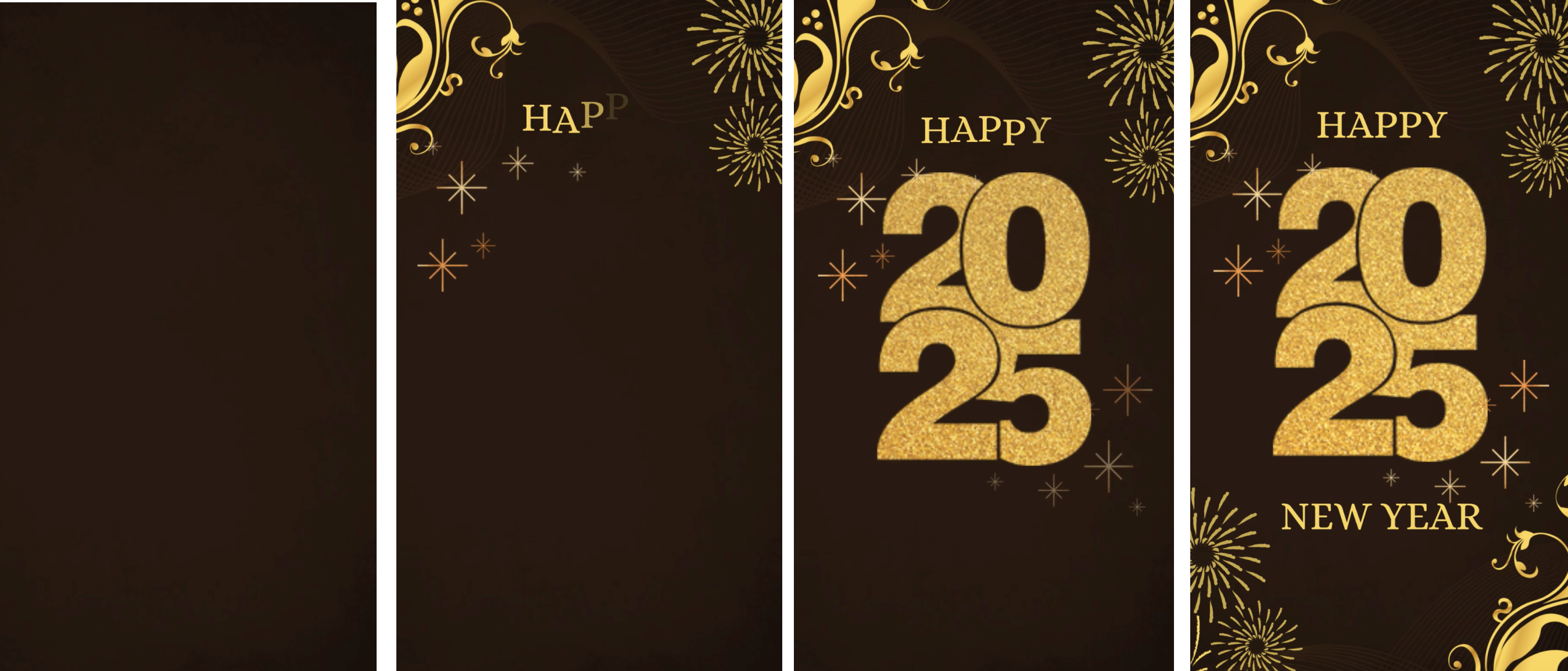}
\caption{Animation of \texttt{Layout 02}
sampled at $t{=}0.0$\,s, $t{=}0.4$\,s (mid-tumble of group~\textsc{0-4},
partial typewriter reveal), $t{=}1.1$\,s (tumble settled, pop complete), and
$t{=}1.65$\,s (typewriter string complete). All three animated components
share $t_{\textrm{from}}{=}0$\,s but differ in animation duration ($1.12$,
$0.56$, $1.65$\,s respectively).}
\label{fig:full-2-storyboard}
\end{figure}

\begin{table}[h!]
\centering
\small
\begin{tabular}{ll}
\toprule
\textbf{Property} & \textbf{Value} \\
\midrule
Canvas             & $1080{\times}1920$, background \texttt{rgb(225,229,234)} \\
Total duration     & 1.5\,s \\
Total components (recursive) & 39 \\
Animated components & 3 (1~\textsc{group}, 1~\textsc{image}, 1~\textsc{text}) \\
Animation primitives & \texttt{tumble}, \texttt{pop}, \texttt{typewriter} \\
Directions invoked   & \texttt{auto} (rotation), \texttt{unknown} (scale, progressive reveal) \\
Entrance regime    & Co-occurring at $t{=}0$, varying durations \\
\bottomrule
\end{tabular}
\caption{Summary of full-layout Example~2.}
\label{tab:full-2-summary}
\end{table}

%--------------------- FULL LAYOUT 3 ----------------------------
\subsubsection{Full Layout 03}
\label{app:full-4}

This 1080$\times$1920 layout exercises the pipeline at the upper end of \emph{animated-component density}: every component in the tree is animated (12 of 12, in contrast to FL1's 7 of 18 and FL2's 3 of 39). It also uniquely combines three observable motion classes within a single composition: \texttt{pop} (2~images plus a \texttt{pop}-class text), \texttt{rotate} (6~tumbling images plus a tumbling banner pair), and the springy entrance family (\texttt{burst}, \texttt{bounce}). All 12 components share $t_{\textrm{from}}{=}0$ but differ in animation duration ($0.56$\,s for \texttt{pop}; $1.12$\,s for \texttt{tumble}; $1.68$\,s for \texttt{burst}; $2.0$\,s for \texttt{bounce}), so the entrance window resolves into four nested cohorts that complete in sequence within the first two seconds. The six-image \texttt{tumble} cluster makes this our most informative test for rotate-class detection under multi-component contention, since any within-class confusion or cross-class leakage to the adjacent \texttt{pop} images directly degrades motion-type accuracy on the tracker-reliable subset. The Track~1 prompt follows the schema given in Example~1.

\begin{figure}[h!]
\centering
\includegraphics[width=0.95\textwidth]{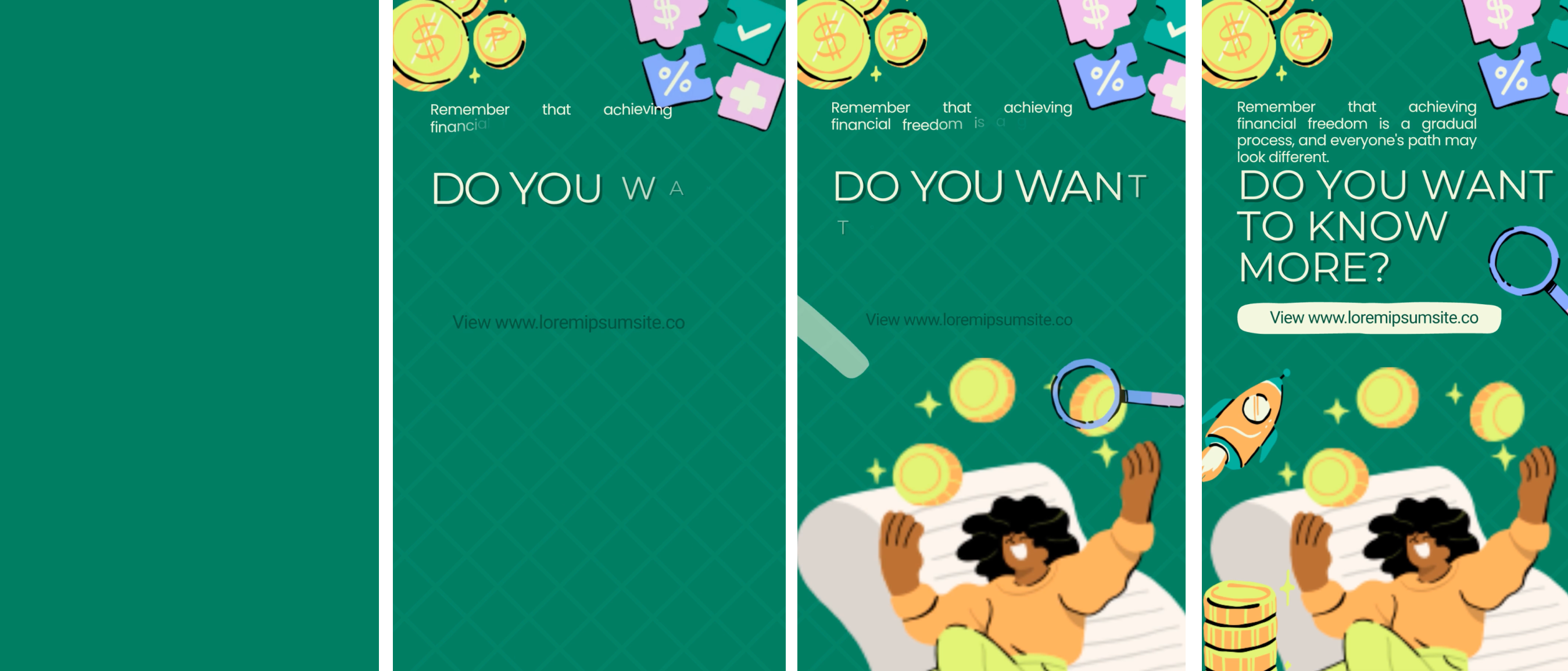}
\caption{Storyboard of \texttt{Full Layout 03} sampled at four points across the entrance window: $t{=}0.0$\,s (initial state), $t{=}0.56$\,s (\texttt{pop} cohort complete: \textsc{0-0}, \textsc{0-3}, \textsc{0-11}), $t{=}1.12$\,s (\texttt{tumble} cohort complete: \textsc{0-1}, \textsc{0-2}, \textsc{0-4}--\textsc{0-8}), and $t{=}2.0$\,s (text \texttt{burst} \textsc{0-9} and \texttt{bounce} \textsc{0-10} complete, near final frame). The 12 animated components together exhaust the component tree.}
\label{fig:full-4-storyboard}
\end{figure}

\begin{table}[h!]
\centering
\small
\begin{tabular}{ll}
\toprule
\textbf{Property} & \textbf{Value} \\
\midrule
Canvas             & $1080{\times}1920$, background \texttt{rgb(1,129,88)} \\
Total duration     & 5.0\,s \\
Total components (recursive) & 12 \\
Animated components & 12 (9~\textsc{image}, 3~\textsc{text}) \\
Animation primitives & \texttt{pop}, \texttt{tumble}, \texttt{burst}, \texttt{bounce} \\
Directions invoked   & \texttt{auto} (rotation), \texttt{unknown} (scale, springy) \\
Entrance regime    & Parallel at $t{=}0$, four duration cohorts \\
                   & ($0.56$, $1.12$, $1.68$, $2.0$\,s) \\
\bottomrule
\end{tabular}
\caption{Summary of full-layout Example~3.}
\label{tab:full-4-summary}
\end{table}

%================================================================
\subsection{Single-Component Examples}
\label{app:single-examples}

Each single-component sample isolates exactly one animated element on a canvas (white default; black when the component colour is
near-white, per Section~3). The eight examples below are organized by the
observable motion class assigned by our normalization map $\varphi$
(Section~4.2). We provide
the verbatim Track~1 prompt for sample~S3 (the \textsc{text} entry, which
exercises the richest field set including font, colour, and line-height
attributes); all other single-component prompts follow the same schema.

\begin{figure}[h!]
\centering
\includegraphics[width=0.95\textwidth]{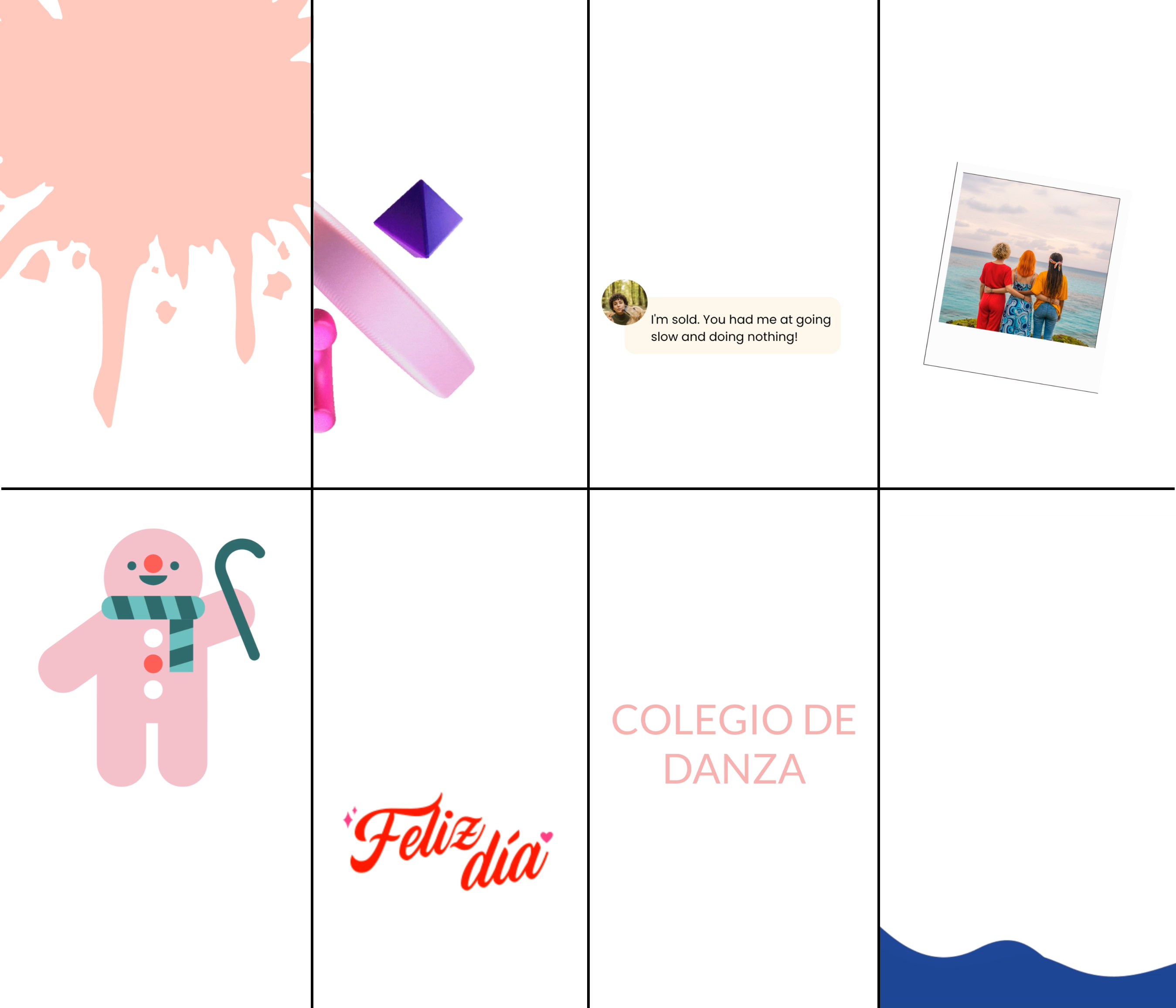}
\caption{Final-frame thumbnails of the eight single-component examples,
grouped by observable motion class. Top row: \texttt{scrapbook} class
(translational entries via LICA \texttt{rise}). Bottom-left pair:
\texttt{rotate} class (\texttt{tumble}). Bottom-right pair: \texttt{pop}
class. Component-type diversity (image, text, group) is balanced across
classes.}
\label{fig:singles-grid}
\end{figure}

\begin{table}[h!]
\centering
\small
\begin{tabular}{cllllrr}
\toprule
\textbf{\#} & \textbf{Class} & \textbf{Type} & \textbf{LICA anim.} & \textbf{Direction} & \textbf{Anim. dur.} & \textbf{Visible until} \\
\midrule
S1 & \texttt{scrapbook} & \textsc{image} & \texttt{rise}   & \texttt{up}         & $0.56$\,s & $10.04$\,s \\
S2 & \texttt{scrapbook} & \textsc{image} & \texttt{rise}   & \texttt{up}         & $0.56$\,s & $8.00$\,s  \\
S3 & \texttt{scrapbook} & \textsc{text}  & \texttt{rise}   & \texttt{up}         & $0.56$\,s & $5.00$\,s  \\
S4 & \texttt{scrapbook} & \textsc{group} & \texttt{rise}   & \texttt{up}         & $0.56$\,s & $5.83$\,s  \\
S5 & \texttt{rotate}    & \textsc{group} & \texttt{tumble} & \texttt{irrelevant} & $1.12$\,s & $1.50$\,s  \\
S6 & \texttt{rotate}    & \textsc{image} & \texttt{tumble} & \texttt{irrelevant} & $1.12$\,s & $5.76$\,s  \\
S7 & \texttt{pop}       & \textsc{image} & \texttt{pop}    & \texttt{irrelevant} & $0.56$\,s & $5.00$\,s  \\
S8 & \texttt{pop}       & \textsc{image} & \texttt{pop}    & \texttt{irrelevant} & $0.56$\,s & $8.00$\,s  \\
\bottomrule
\end{tabular}
\caption{Single-component samples with observable motion class, LICA
animation primitive, direction, animation duration, and component visible
duration. \texttt{irrelevant} denotes a non-directional motion
(\texttt{tumble} and \texttt{pop} have no semantic compass direction).}
\label{tab:singles-summary}
\end{table}

\paragraph{Track~1 prompt for S3 (verbatim).}
S3 is a \texttt{rise up} \textsc{text} entry rendering the headline
``COLEGIO DE DANZA'' on a white canvas. We show the prompt verbatim to
illustrate the full text-component field schema (\texttt{Content},
\texttt{Font}, \texttt{Color}, line-height, letter-spacing, text alignment),
which is a strict superset of the \textsc{image} and \textsc{group} field
sets used by the other seven samples.
 
{\scriptsize
\begin{Verbatim}[breaklines=true,breakanywhere=true]
Generate a 4.75s animated video at 1080x1920 pixels with background color rgb(255, 255, 255).
 
Animated components: 1 (static components may also be present for context but are not listed below).
 
Component 1: TEXT
- Content: "COLEGIO DE DANZA"
- Font: Lato--400, size 133px, weight 400, style normal
- Color: rgb(239, 161, 168)
- Line height: 159.0px, letter spacing: 0em, text alignment: center
- Position: left=108px, top=671.777px
- Size: width=864px, height=317.986px
- Motion type: rise
- Direction: up
- Animation duration: 0.56s
- Speed: 1
- Trigger: both (begins at t=0s)
- Remains visible until t=5.0s
 
Keep static components visible as context; only the listed animated components should be described for motion.
\end{Verbatim}
}

%__________________________________________________________________
\newpage
\section{Motion-Type Classifier: Detailed Rules}
\label{app:motion_rules}

This appendix expands the high-level rule cascade from Table~\ref{tab:motion_rules} (Section~\ref{sec:framework_motion_type}) into the exact thresholds and per-rule rationale used by the released implementation. The $12$ rules in the main paper map onto $15$ rules here because two collapse two cases each: the opacity-transient rule splits into a contour-tracker variant and a YOLO-tracker variant, and the late opacity/presence-change rule splits on residual energy.

\subsection{Feature definitions}

For each tracked component we derive five 1-D signals: centroid trajectory $(x_t, y_t)$, scale curve $s_t = \sqrt{(w_t / \tilde{w})(h_t / \tilde{h})}$, OBB rotation $\theta_t$, opacity proxy $o_t$ (detector confidence in YOLO mode; bbox area normalized by its running peak in contour mode), and presence $p_t\in\{0,1\}$. Scalar features used by the classifier:

\begin{itemize}\itemsep1pt
    \item $d_{\text{net}}$: net centroid displacement, normalized by the canvas diagonal. $\Delta_{\max}$: maximum per-frame step.
    \item $\text{scale range}$, $\text{pop\_height}$ (largest scale spike above the settled baseline), $\text{start\_extent}$ ($\Delta_{\max}$ over the first quarter of the clip).
    \item $E_{\text{start/mid/end}}$: composite motion energy summed over the three thirds of the clip; $E_{\text{peak}}=\max(E_{\text{start}}, E_{\text{end}})$, $\rho=E_{\text{mid}}/E_{\text{peak}}$, $E_{\text{tot}}=\sum E$.
    \item $\theta_{\text{tot}}$, $\theta_{90}$ (period-$90^{\circ}$ unwrap, resolving the axis-flip ambiguity of \texttt{cv2.minAreaRect}), $\dot\theta_{\max}$.
    \item $\text{pos\_zx}, \text{sc\_zx}$ (zero crossings of the detrended centroid and scale curves).
    \item $\text{op range}$, $\text{op}_{\text{start/mid/end}}$, $\text{low\_mid\_frac}$ (fraction of mid-third frames with opacity proxy below $0.5$; YOLO-mode only).
    \item $\text{pres}_{\text{start/mid/end}}$ (mean of $p_t$ over each third).
\end{itemize}

All length features are normalized by the canvas diagonal. We use $\varepsilon = 3\times 10^{-3}$ as the positional noise floor across both backends.

\subsection{Rule cascade}

\begin{table}[h]
\centering
\caption{Detailed motion-type rules with full thresholds. Rules are evaluated top-down; the first match wins.}
\label{tab:motion_rules_full}
\footnotesize
\resizebox{\columnwidth}{!}{
\begin{tabular}{@{}cll@{}}
\toprule
\textbf{\#} & \textbf{Condition} & \textbf{Label} \\
\midrule
\midrule
1  & $\text{pres}_{\text{full}}<0.05$ & \textit{static} \\
\midrule
\multirow{2}{*}{2}  & contour mode; $\text{op range}>0.30$, $d_{\text{net}}<0.06$, and either $\text{op}_{\text{mid}}-\min(\text{op}_{\text{start}},\text{op}_{\text{end}})>0.15$ or & \multirow{2}{*}{\textit{fade}} \\
& $\min(\text{pres}_{\text{start}},\text{pres}_{\text{end}})<0.7$ with $\text{pres}_{\text{mid}}>0.85$ and the other endpoint $>0.85$ & \\
\midrule
3  & YOLO mode; $\text{low\_mid\_frac}>0.85$, $\text{pop\_height}<0.20$, $\text{scale range}<0.50$, $d_{\text{net}}<0.05$, $E_{\text{mid}}<0.05$ & \textit{fade} \\
\midrule
4  & $E_{\text{peak}}>5\times 10^{-3}$, $\rho<0.4$, $\text{pop\_height}>0.10$, $\text{start\_extent}<0.005$, $d_{\text{net}}<0.03$, $E_{\text{mid}}<0.05$ & \textit{pop} \\
\midrule
5  & $E_{\text{peak}}>5\times 10^{-3}$, $\rho<0.4$, $|\theta_{90}|\geq 90^{\circ}$ & \textit{rotate} \\
\midrule
6  & $E_{\text{peak}}>5\times 10^{-3}$, $\rho<0.4$, otherwise & \textit{scrapbook} \\
\midrule
7  & $|\theta_{90}|>45^{\circ}$, $\text{scale range}<0.20$, $d_{\text{net}}<0.05$ & \textit{rotate} \\
\midrule
8  & $\big(\text{pos\_zx}\geq 6,\,\Delta_{\max}>0.05\big)$ \emph{or} $\big(|\theta_{\text{tot}}|<30^{\circ},\,\dot\theta_{\max}>100^{\circ}/\text{s}\big)$; $d_{\text{net}}<0.03$ & \textit{wiggle} \\
\midrule
9  & $\text{sc\_zx}\geq 6$, $\text{scale range}>0.06$, $\Delta_{\max}<0.005$ & \textit{breathe} \\
\midrule
10 & $d_{\text{net}}>0.10$, $E_{\text{mid}}>0.02$, $d_{\text{net}}/E_{\text{tot}}>0.4$ & \textit{pan} \\
\midrule
11 & $\text{op range}>0.30$, $\text{scale range}<0.15$, $\Delta_{\max}<2\varepsilon$, $d_{\text{net}}<2\varepsilon$ & \textit{fade} \\
\midrule
12 & $\min(\text{pres}_{\text{start}},\text{pres}_{\text{end}})<0.5$, $\text{pres}_{\text{mid}}>0.85$, $E_{\text{tot}}>0.05$ & \textit{scrapbook} \\
\midrule
13 & $\min(\text{pres}_{\text{start}},\text{pres}_{\text{end}})<0.5$, $\text{pres}_{\text{mid}}>0.85$, $E_{\text{tot}}\leq 0.05$ & \textit{fade} \\
\midrule
14 & $E_{\text{tot}}>0.05$ & \textit{scrapbook} \\
\midrule
15 & none of the above & \textit{static} \\
\bottomrule
\end{tabular}
}
\end{table}

\paragraph{Rationale.} Rule~1 guards against degenerate chains. Rules~2--3 separate fade detection by tracker backend: the contour proxy responds linearly to alpha, so a fade produces both an opacity ramp and contour-shape jitter; YOLO confidence drops on \emph{any} hard-to-localise frame, so the YOLO rule conditions on a sustained low-confidence middle plus the absence of a scale spike or centroid drift. Rules~4--6 implement the LICA on-enter signature -- ``endpoints hot, middle cold'' -- and dispatch to \textit{pop} when the chain is dominated by a scale overshoot, to \textit{rotate} when the period-$90^{\circ}$ unwrapped angle exceeds a quarter-turn (LICA's \textit{tumble}/\textit{roll}), and to \textit{scrapbook} otherwise. Rule~7 catches sustained rotations missed by the transient gate. Rule~8 admits two wiggle signatures: oscillating centroid (large-amplitude back-and-forth, gated to reject YOLO bbox jitter) or oscillating rotation. Rule~9 is the scale analogue. Rule~10 requires a pan to satisfy three conditions simultaneously (large net displacement, sustained mid-energy, high directional fraction $d_{\text{net}}/E_{\text{tot}}$); the directional fraction distinguishes a true pan from any oscillating motion that happens to travel. Rule~11 catches fades whose tracker signature was too weak to fire rules~2--3. Rules~12--13 are the presence-rescue branch for chains where the tracker lost the component during the entry/exit transient itself; we split on residual energy because in the low-energy bucket both true fades and true scrapbook entries are admissible, and \textit{fade} is the higher-expected-value choice given the partial-credit table ($5$ fade GT vs.\ $4$ scrapbook$+$pop GT in this bucket on the GT data). Rule~14 catches slow continuous motion below the pan threshold.

\subsection{Partial-credit scoring}

\begin{table}[h]
\centering
\small
\caption{Partial-credit table for motion-type predictions. Rows are GT, columns are predicted; blank cells score $0$. The table is intentionally asymmetric on rotate/scrapbook and pan/scrapbook.}
\label{tab:partial_credit}
\begin{tabular}{l|cccccc}
\toprule
\textbf{GT $\downarrow$ \,/\, Pred $\rightarrow$} & \textit{scrap.} & \textit{pop} & \textit{fade} & \textit{rotate} & \textit{pan} & \textit{wig./bre.} \\
\midrule
\textit{pop}       & $0.5$ & $1.0$ & --    & --    & --    & --    \\
\textit{scrapbook} & $1.0$ & $0.5$ & $0.3$ & $0.5$ & $0.3$ & --    \\
\textit{fade}      & $0.3$ & --    & $1.0$ & --    & --    & --    \\
\textit{rotate}    & $1.0$ & --    & --    & $1.0$ & --    & --    \\
\textit{wiggle}    & --    & --    & --    & --    & --    & $0.4$ \\
\textit{breathe}   & --    & --    & --    & --    & --    & $0.4$ \\
\bottomrule
\end{tabular}
\end{table}

\textit{Rotate $\to$ scrapbook} earns full credit because LICA's \textit{tumble}/\textit{roll} entries combine a translational entry with rotation, so a scrapbook prediction is observationally equivalent to the OBB signal. \textit{Scrapbook $\to$ pan} is asymmetric (only credited in one direction): a scrapbook that travels far enough to satisfy the pan gate is observationally close to a pan, but the reverse miss indicates the classifier inferred a transient on a sustained motion and is counted as a hard miss. On the single-component track LICA releases no motion label for some prompts (carried as \textit{unknown}); the metric awards $1.0$ for predictions of \textit{static} or \textit{unknown} on these and $0$ otherwise, turning the score into a spurious-motion fraction.

\subsection{Threshold derivation}

All thresholds were chosen by inspecting GT-render feature distributions, not learned from generated videos. Four are most consequential. (i)~$\varepsilon=3\times 10^{-3}$: upper edge of contour-tracker centroid jitter on static-class GT renders, padded for YOLO mode. (ii)~$E_{\text{peak}}>5\times 10^{-3}$: chosen so LICA's shortest authored animations (\textit{photoRise}, $\sim 0.42$\,s) are admitted while contour jitter on static GT renders is rejected. (iii)~$\rho<0.4$: transient chains cluster at $\rho<0.2$ on GT, sustained chains at $\rho>0.7$, and $0.4$ leaves a clean margin in both directions. (iv)~$E_{\text{tot}}>0.05$: empirical valley between the static-GT mode ($\approx 0.01$) and the slow-scrapbook GT mode ($\approx 0.2$). The rest follow the same recipe: the empirical valley between the targeted class and its nearest confusable class.

%__________________________________________________________________
\clearpage
\section{Motion-Type Classification Results}
\label{app:confusion-matrices}

\begin{figure}[h]
    \centering
    \includegraphics[width=14cm]{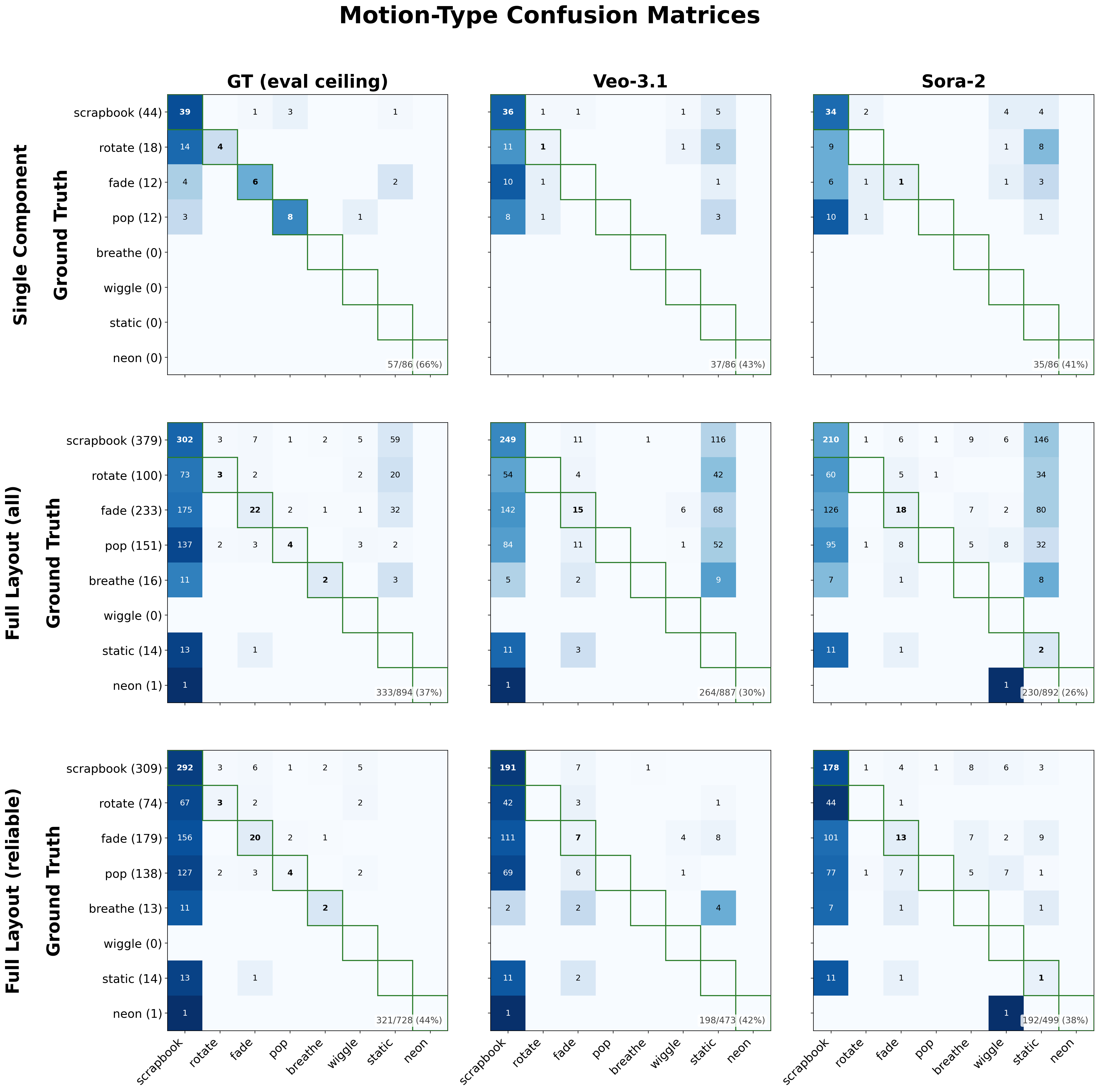}
    \caption{Motion-type confusion matrices for GT renders, Veo-3.1, and Sora-2 across the single-component track (top), the full-layout track with all components (middle), and the tracker-reliable subset of the full-layout track (bottom). On single-component GT, scrapbook is strongly diagonal (89\%), but the classifier collapses most rotate-family samples into scrapbook (14/18); fade and pop show moderate leakage to scrapbook as well. On the full-layout tracks the dominant pattern intensifies: fade, pop, and rotate rows collapse almost entirely to the scrapbook column, because the YOLO-OBB tracker cannot resolve the opacity signal that distinguishes fades nor the brief scale spike that distinguishes pops from a generic translational entry. Rotate-family components predicted as scrapbook receive full credit in the partial-credit scoring (Table~\ref{tab:motion_rules}) and are not penalized. Empty rows (e.g.\ wiggle in full-layout; breathe, wiggle, static, neon in single-component) indicate motion types absent from the ground-truth distribution at that track.}
    \label{fig:confusion-matrices}
\end{figure}

%__________________________________________________________________
\newpage
\section{LLM Text Evaluator Prompt}
\label{app:llm-text-prompt}
For LLM-based text recoverability evaluation, each sampled video frame is encoded as a JPEG image and passed to the model with a strict OCR instruction. The evaluator is asked to transcribe only visible text and return a JSON object containing one string per visual text line. We use the same prompt template across LLM evaluators unless otherwise noted.
\paragraph{System prompt.}
\begin{quote}
\small
\ttfamily
You are a strict OCR transcription engine for one video frame.\\
Transcribe exactly what is visible, without interpretation.\\
Rules:\\
1) Do not translate, paraphrase, summarize, or grammar-correct.\\
2) Preserve exact characters: case, digits, punctuation, symbols, accents, apostrophes.\\
3) Preserve visual reading order: top-to-bottom, then left-to-right.\\
4) Keep each visual text line as a separate line item.\\
5) Never reorder words inside a line.\\
6) Never hallucinate missing words; if a character is unreadable, use '?'.\\
7) If no readable text exists, return an empty lines list.\\
Return only a JSON object with this shape: \{"lines": ["line1", "line2"]\}.\\
No markdown fences. No extra keys. No explanations.
\end{quote}
\paragraph{User payload.}
\begin{quote}
\small
\ttfamily
\{\\
\quad "prompt\_version": "default",\\
\quad "task": "ocr\_from\_single\_video\_frame",\\
\quad "language\_hint": "en",\\
\quad "return\_schema": \{ "lines": ["string"] \}\\
\}
\end{quote}
The image is provided together with the user payload. The returned \texttt{lines} are parsed as text candidates and then evaluated using the text recoverability metric in Section~\ref{sec:eval_framework}.

%__________________________________________________________________
\section{Text Recoverability Backend Validation}
\label{app:text-backend}

We compare OCR and LLM-based evaluators for the text recoverability metric. Prior work on aesthetic poster generation notes that traditional OCR can be unreliable for artistic typography and stylized design text, and uses VLM-based recognition to handle text regions in poster layouts~\cite{chen2025postercraft}. Motivated by this observation, we evaluate both conventional OCR backends and LLM-based transcription backends. In all cases, the backend is used only to extract text candidates from sampled frames; the final score is computed with the same CER-based assignment metric defined in Section~\ref{sec:eval_framework}.

Full-layout runs use the video set with text-containing samples. All backends are evaluated on the same frames sampled at 2 fps, following the frame-sampling ablation in Appendix~\ref{app:frame-sampling}. The single-component rows provide a controlled reference setting. For single-component LLM evaluation, we use Claude Opus 4.7 only, selected based on the full-layout backend comparison.

\begin{table}[h]
\centering
\small
\begin{tabular}{llcc}
\toprule
Dataset & Backend & Best AR \(\uparrow\) & Hard AR \(\uparrow\) \\
\midrule
Single-component & EasyOCR & 0.919 & 0.769 \\
Single-component & PP-OCRv5 & 0.769 & 0.769 \\
Single-component & Tesseract & 0.695 & 0.462 \\
Single-component & Claude Opus 4.7 & \textbf{1.000} & \textbf{1.000} \\
\midrule
Full-layout & PP-OCRv5 & 0.862 & 0.659 \\
Full-layout & EasyOCR & 0.903 & 0.584 \\
Full-layout & Tesseract & 0.565 & 0.314 \\
Full-layout & GPT-5.5 & 0.951 & 0.920 \\
Full-layout & Gemini 3.1 Pro & 0.950 & 0.922 \\
Full-layout & Claude Opus 4.7 & \textbf{0.960} & \textbf{0.940} \\
\bottomrule
\end{tabular}
\caption{Backend validation for text recoverability. Best AR is the average per-component best-over-time recognition score, and Hard AR is the fraction of required text components exactly recovered. \textbf{Bold} indicates the best backend per dataset.}
\label{tab:text-backend-validation}
\end{table}

%__________________________________________________________________
\section{Frame Sampling Rate Ablation}
\label{app:frame-sampling}

We evaluate the effect of frame sampling rate on text recoverability using the full-layout validation set. We vary the sampling rate from 0.5 to 8 fps while keeping the evaluator and assignment metric fixed. This ablation is used to justify the 2 fps default used in the main experiments.

\begin{figure}[h]
\centering
\includegraphics[width=0.95\textwidth]{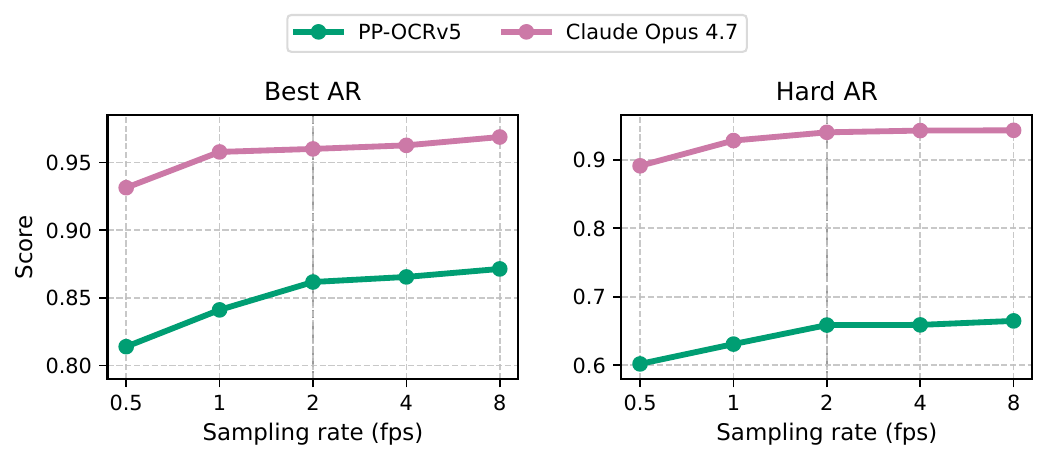}
\caption{Effect of frame sampling rate on text recoverability in full-layout validation scenes. Both evaluators improve rapidly up to 2 fps and show diminishing returns beyond that point, supporting the 2 fps default used in the main experiments.}

\label{fig:frame-sampling-ablation}
\end{figure}

\end{document}